\documentclass[onecolumn,amsmath,amssymb,notitlepage,numbers,superscriptaddress,noeprint]{revtex4-1}

\usepackage{graphicx}
\graphicspath{
      {figures/}
      {../figures/}
    }
\usepackage{dcolumn}
\usepackage{bm}
\usepackage{float}
\usepackage{physics}

\begin{document}

\title{A scalable gallium-phosphide-on-diamond spin-photon interface}

\author{Nicholas S. Yama}
\affiliation{University of Washington, Electrical and Computer Engineering Department, Seattle, WA, 98105, USA}
\author{Chun-Chi Wu}
\affiliation{University of Washington, Electrical and Computer Engineering Department, Seattle, WA, 98105, USA}
\author{Fariba Hatami}
\affiliation{Department of Physics, Humboldt-Universitat zu Berlin, Newtonstrasse, Berlin, 10117, Germany}
\author{Kai-Mei C. Fu}
\affiliation{University of Washington, Electrical and Computer Engineering Department, Seattle, WA, 98105, USA}
\affiliation{University of Washington, Physics Department, Seattle, WA, 98105, USA}
\affiliation{Physical Sciences Division, Pacific Northwest National Laboratory, Richland, Washington 99352, USA}

\begin{abstract}
    The efficient interfacing of quantum emitters and photons is fundamental to quantum networking.
    Quantum defects embedded in integrated nanophotonic circuits are promising for such applications due to the deterministic light-matter interactions of high-cooperativity ($C>1$) cavity quantum electrodynamics and potential for scalable integration with active photonic processing.
    Silicon-vacancy (SiV) centers embedded in diamond nanophotonic cavities are a leading approach due to their excellent optical and spin coherence, however their long-term scalability is limited by the diamond itself, as its suspended geometry and weak nonlinearity necessitates coupling to a second processing chip.
    Here we realize the first high-cooperativity coupling of quantum defects to hybrid-integrated nanophotonics in a scalable, planar platform.
    We integrate more than 600 gallium phosphide (GaP) nanophotonic cavities on a diamond substrate with near-surface SiV centers.
    We examine a particular device with two strongly coupled SiV centers in detail, confirming above-unity cooperativity via multiple independent measurements.
    Application of an external magnetic field via a permanent magnet enables optical resolution of the SiV spin transitions from which we determine a spin-relaxation time $T_1>0.4$\,ms at 4\,K.
    We utilize the high cooperativity coupling to observe spin-dependent transmission switching and the quantum jumps of the SiV spin via single-shot readout.
    These results, coupled with GaP's strong nonlinear properties, establish GaP-on-diamond as a scalable planar platform for quantum network applications.
\end{abstract}

\maketitle
\vspace{-2em} 

\section{Introduction}
The efficient coupling of optically active spin defects in diamond to nanophotonic cavities has enabled important advancements in the realization of cavity quantum electrodynamics (cQED)-based quantum networks~\cite{stas2022sivregister,bersin2024sivnetworking,knaut2024sivnetwork}.
However, a major obstacle to the long-term scalability of such technologies is the underlying design paradigm, in which the photonic cavity is formed directly in suspended~\cite{burek2014diamondcavities,sipahigil2016phc,nguyen2019sivregister} or thin-film~\cite{ding2024diamondmembrane,riedel2025lightsynq} diamond structures.
Such fabrication processes are not only difficult to perform, but may introduce surface defects which diminish the defect's optical and spin coherence~\cite{cui2015plasmadamage,chakravarthi2021nvsurface} (especially for non-centrosymmetric defects such as the nitrogen-vacancy (NV) center~\cite{sipahigil2014indistinguishable}).
Moreover, diamond lacks the optical nonlinearities necessary for realizing active photonic elements --- a key requirement of scalable quantum photonic processing.
Consequently, the long-term scaling of such devices will require the transfer of cavities onto more suitable photonic circuit platforms~\cite{wan2020large,riedel2023linbo3,li2024cmos}.
The necessity of bespoke pick-and-place fabrication, diamond etching, and device suspension, however, remain significant challenges to scalability.

Hybrid-integrated gallium phosphide (GaP)-on-diamond photonics have been proposed as a scalable alternative.
Here, the photonic circuit is instead etched into a thin-film GaP layer bonded onto the diamond substrate in a planar geometry~\cite{barclay2009gapphc,gould2015gap} without the need for diamond etching~\cite{chakravarthi2023stamp}.
GaP's unique combination of a large refractive index ($3.21$ at $737$\,nm) and band gap ($2.24$\,eV) enables the low-loss guiding of spin defect photoluminescence.
GaP also has a large bulk second-order nonlinearity ($\chi^{(2)} = 110$\,pm/V~\cite{corso1996gapChi2}), enabling efficient nonlinear devices~\cite{wilson2020nonlineargap,honl2022gaptransduction,logan2023triply}.
Moreover, its high speed-of-sound contrast with diamond and large acousto-optic coefficient ($p_{11} = -0.24$~\cite{mytsyk2015gapphotoelastic}) makes GaP-on-diamond a promising platform for integrated phononics and cavity-optomechanics~\cite{schneider2019gapoptomech,xinyuan2023gapoptomech}.
Finally, GaP can be lattice matched to silicon for large-scale thin-film growth, enabling a straightforward means of long-term scalability~\cite{yama2024bgap}.
However, the intrinsic separation of the spin defect and cavity mode (imposed by the hybrid geometry) reduces the interaction strength, historically limiting experimental cQED cooperativities to $C\approx0.1$ in both diamond~\cite{chakravarthi2023stamp} and rare-earth-ion~\cite{dibos2018hybrid,ourari2023hybrid,wu2023hybrid} devices.
The attainment of deterministic interactions in the high-cooperativity regime ($C>1$) thus remains an important milestone in the development of scalable quantum networks.

In this work we address this challenge through a modified design process which yields a nearly 100-fold improvement in simulated performance compared to prior hybrid designs~\cite{chakravarthi2023stamp} and is comparable to existing all-diamond devices.
We demonstrate the scalable and high-yield fabrication of these devices and realize the first high-cooperativity coupling of silicon-vacancy (SiV) centers in a hybrid-integrated platform.
We then demonstrate important functionalities of the spin-photon interface, including spin-dependent optical transmission switching and efficient single-shot readout of the SiV spin state.
Our results demonstrate the potential for GaP-on-diamond-based quantum networking technologies with long-term scalability.

\section{GaP-on-diamond cavity design}
\begin{figure}[t]
    \centering
    \includegraphics{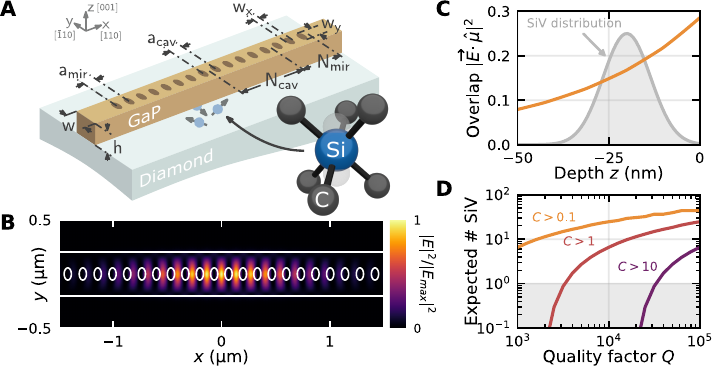}
    \caption{
        \textbf{Cavity design.}
        (\textbf{A}) Schematic of the GaP-on-diamond cavity design.
        (\textbf{B}) Simulated field profile of the fundamental mode at the center of the nanobeam.
        (\textbf{C}) Simulated cavity mode overlap versus $z$ at the center of the cavity for an SiV center oriented along a perpendicular $\langle111\rangle$ axis.
        The gray Gaussian curve indicates the estimated implanted SiV distribution with mean depth $\ev{z}=-20.0$\,nm and straggle $\sigma=6.5$\,nm.
        (\textbf{D}) The expected number of coupled SiV with $C>0.1$ (orange), $1$ (red), and $10$ (purple) versus total quality factor $Q$.
        The calculation assumes an SiV density of $50$\,\textmu m$^{-2}$ and an intrinsic radiative efficiency $\gamma_0/\gamma = 0.07$~\cite{chakravarthi2023stamp}.
    }
    \label{cavity-design}
\end{figure}

The cavity is formed from a 1-d GaP photonic crystal (PhC) nanobeam heterogeneously integrated on a planar diamond substrate with near-surface, negatively charged SiV centers (Fig.~\ref{cavity-design}A).
The nanobeam cavity consists of a GaP waveguide of width $w$ and height $h$ with a series of identical elliptical holes (minor/major axes given by $w_x$ and $w_y$) symmetrically arranged along the waveguide length.
The lattice constant $a$ is symmetrically chirped from $a_{\mathrm{cav}}$ to $a_{\mathrm{mir}}$ over $N_{\mathrm{cav}}$ unit cells in either direction from the center.
An additional $N_{\mathrm{mir}}$ cells with spacing $a_{\mathrm{mir}}$ are included on both ends.

The primary design objective is to maximize the cooperativity
\begin{equation}
    \label{eq:cooperativity}
    C =
    \frac{3}{4\pi} 
    \bigg(\frac{\gamma_0}{\gamma}\bigg)
    \bigg(\frac{Q}{V}\bigg) 
    \bigg( \frac{\bm\mu}{|\bm\mu|} \cdot \frac{\vb E(\vb x_a)}{\max|\vb E|} \bigg)^2,
\end{equation}
where $(\gamma_0/\gamma)$ denotes the intrinsic relative decay rate of the target transition, $Q$ is the quality factor and $V$ is the wavelength-normalized mode volume.
The last factor describes the geometric ``overlap'' of the SiV transition dipole moment $\bm\mu$ (at position $\vb x_a$) with the cavity mode field $\vb E(\vb x)$.
With respect to the optimization of all-diamond structures~\cite{nguyen2019sivregister}, the overlap factor is often neglected as it is saturated by ideal SiV placement.
In contrast, for hybrid devices, the overlap factor cannot be maximized without adversely affecting $Q/V$, and so it is instead an important design metric in and of itself.
Consequently, known PhC design principles which strictly maximize $Q/V$~\cite{quan2010phcdesign,quan2011phcdesign} are insufficient.
Instead we perform a two-stage design optimization process consisting of a forward-design process to maximize $Q/V$, followed by an inverse-design parameter search to maximize the cooperativity (Supplemental 1B).

The final optimal design is given by parameters $(w,h,w_x,w_y,a_{\mathrm{cav}},a_{\mathrm{mir}}) = (402.8, 182.8, 66.3, 112.0, 132.5, 140.1)$\,nm and $N_\mathrm{cav} = 12$.
Fig.~\ref{cavity-design}B illustrates the simulated cavity mode at the center of the GaP beam which has an intrinsic, radiation-limited quality factor of $Q_{i,\mathrm{sim}} = 3\times 10^5$ and a normalized mode volume of $V=2.0(\lambda/n)^3$.
The simulated overlap factor is plotted in Fig.~\ref{cavity-design}C, attaining a value of $0.17$ for a $\langle111\rangle$-oriented dipole at the mean implantation depth of $20$\,nm.
Fig.~\ref{cavity-design}D shows the the expected number of SiV to attain a given cooperativity versus $Q$ in our sample.
For a cavity with total loaded quality factor $Q>3200$, we expect there to be at least one coupled SiV with $C>1$ on average.
The resulting design represents a nearly 100-fold increase in simulated performance compared to Ref.~\cite{chakravarthi2023stamp}.

\section{Large-scale integration}
More than 4000 PhC devices are fabricated on a 200-nm GaP membrane on AlGaP sacrificial layer using a standard electron-beam lithography (EBL) and reactive-ion plasma process, and then suspended by selectively etching the AlGaP layer in hydrofluoric acid.
The EBL write parameters and cavity geometry are varied over the chip to ensure some devices attain resonances in the desired wavelength range (Supplemental 3F).
Transmission measurements of the suspended devices are performed to characterize the resonance wavelength and quality factor $Q$.
We then select 16 sets of devices with resonances near the target wavelength to transfer to a 2-mm diamond chip with implanted SiV centers via a standard dry stamping technique.
The stamped PhC cavities are aligned along a $\langle110\rangle$ edge of the diamond, ensuring that two of the four $\langle111\rangle$ SiV orientations couple maximally to the cavity TE mode.
We successfully transfer $637$ of $640$ devices onto the diamond corresponding to a $99.5$\% yield (Fig.~\ref{large-scale-integration}A).
Complete details of the fabrication process are provided in Supplemental 1C.

\begin{figure}[tp]
    \centering
    \includegraphics[width=\textwidth]{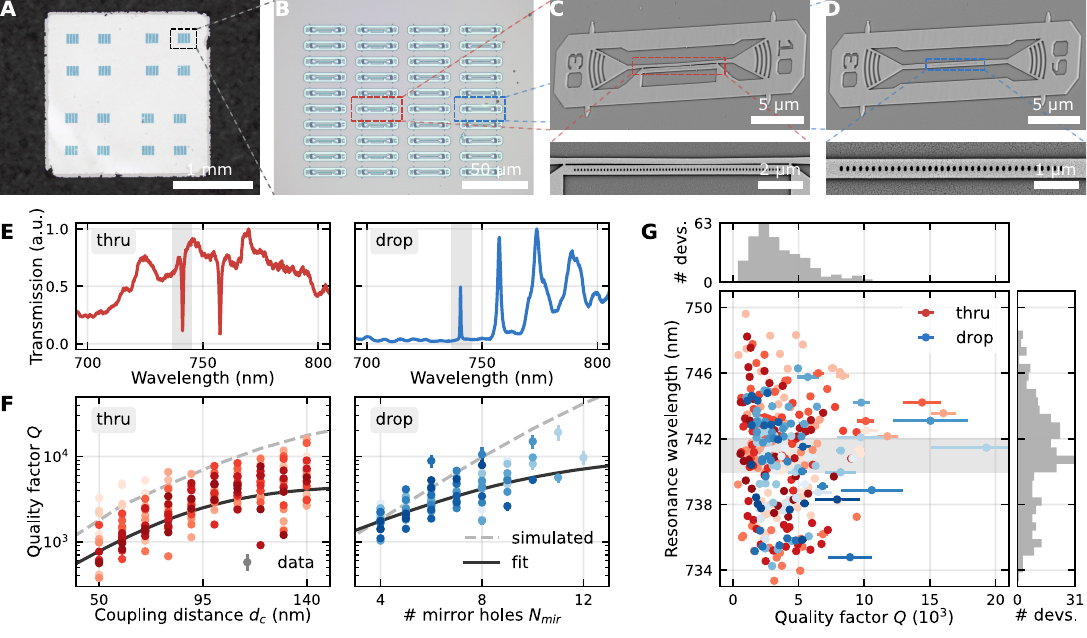}
    \caption{
        \textbf{Large-scale device integration.}
        Optical images of (\textbf{A}) the 2-mm diamond chip and (\textbf{B}) a close up of one of the 16 sets of devices.
        Thru/drop devices are indicated by the red/blue boxes.
        SEM images of the (\textbf{C}) thru and (\textbf{D}) drop devices.
        Texture is due to a gold/palladium coating for imaging.
        (\textbf{E}) Representative transmission spectra at room temperature.
        The primary resonance is highlighted in gray.
        (\textbf{F}) Quality factor versus coupling coupling parameter for all measured devices ($N=321$).
        The color brightness denotes the device set.
        (\textbf{G}) Distribution of the quality factors and resonance wavelengths at room temperature.
        Devices within the gray band between $740$--$742$\,nm can be tuned onto resonance with SiV centers at cryogenic temperatures.
    }
    \label{large-scale-integration}
\end{figure}

Each set of 40 devices consists of 20 side-coupled ``thru'' devices (Fig.~\ref{large-scale-integration}C) and 20 mirror-coupled ``drop'' devices (Fig.~\ref{large-scale-integration}D).
For each type, the waveguide coupling rate is modulated to ensure some fraction of the devices are over coupled.
For the thru devices, the number of mirror holes is fixed at $N_\mathrm{mir}=25$ to eliminate coupling to the PhC waveguide, and the separation of the cavity and coupling waveguide $d_c$ is modulated between 50--140\,nm.
In contrast, for the drop devices, the number of mirror holes $N_\mathrm{mir}$ is modulated between 4--12 and no coupling waveguide is present.

Transmission spectra are obtained by exciting the devices with a broadband supercontinuum laser through one of the free-space elliptical grating couplers and collecting the spatially filtered transmitted light through the other.
Representative transmission spectra for the thru/drop devices are shown in Fig.~\ref{large-scale-integration}E revealing a sharp resonance dip/peak respectively.
Fitting to a Lorentzian enables determination of the resonance position $\omega_0$, total cavity loss rate $\kappa$, and quality factor $Q=\omega_0/\kappa$ (Supplemental 3C).
The quality factor is plotted against the coupling parameter $d_c$, $N_\mathrm{cav}$ for all measured devices ($N=321$) in Fig.~\ref{large-scale-integration}F.
We observe, on average, an exponential increase in $Q$ with the coupling parameter in agreement with simulation.
The measured $Q$ factors saturate below the simulated intrinsic value $Q_{i,\mathrm{sim}}$ as a result of fabrication imperfections.
We observe a number of under-coupled devices with $Q>10^4$, while fitting reveals an average fabrication-limited quality factor of $4670\pm 40$ and $10400\pm 1000$ for the thru and drop devices respectively.
Much of this variation can be attributed to the strong sensitivity of the PhC resonance to slight perturbations of the design parameters, both systematic and random.
This is seen in Fig.~\ref{large-scale-integration}G, which plots the room-temperature resonance wavelength versus quality factor for all measured devices, showing a large distribution of the cavity resonance wavelengths.
The gray region from 740--742\,nm indicates the devices which can be tuned onto resonance at cryogenic temperatures in our setup.

\section{High-cooperativity coupling to SiV centers}
\begin{figure}[t]
    \centering
    \includegraphics{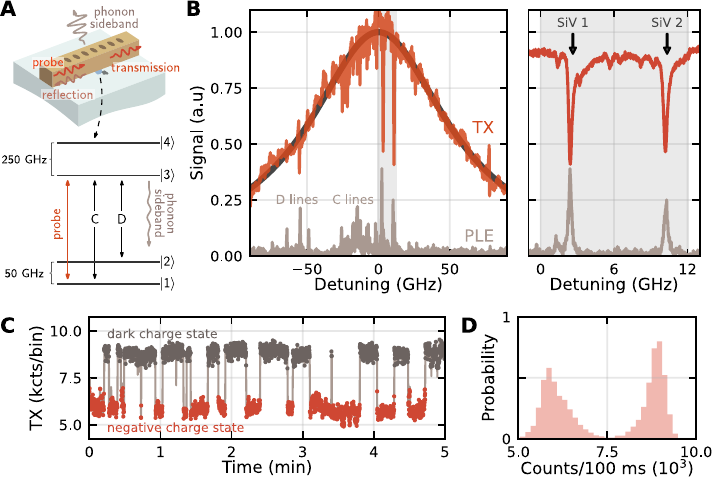}
    \caption{
        \textbf{Dipole-induced scattering.}
        (\textbf{A}) Illustration of the transmission measurement in a drop device.
        (\textbf{B}) Transmission and PLE spectra when in near resonance with the $C$ lines.
        A strong modulation of the transmission is observed for two blue-shifted SiV centers (right: expanded view).
        (\textbf{C}) Single-shot readout of the SiV charge state by resonant scattering off the DIT peak.
        Counts are collected in 100-ms bins.
        (\textbf{D}) Intensity distribution of the single-shot readout.
        Vibrations cause additional broadening of the distribution beyond shot noise.
    }
    \label{dit}
\end{figure}

The sample is cooled to 4\,K to characterize the SiV-cavity coupling.
The cavity resonances blue shift by approximately 5\,nm and can be controllably red-shifted \textit{in situ} by injecting small amounts of xenon gas which condenses onto the sample (Supplemental 3B).
We scan a highly attenuated narrowband diode laser over the individual cavity resonances as they are tuned over the $C$ transition of the SiV zero-phonon line (ZPL) near 737\,nm (Fig.~\ref{dit}A).
The laser excites the devices through one of the free-space grating couplers and the spatially-filtered transmission (TX) can be collected via the other grating coupler.
Alternatively, we collect spectrally filtered phonon-sideband (PSB) emission radiated from the top of the cavity to perform photoluminescence excitation (PLE) spectroscopy.
A 532-nm repump pulse is utilized at the start of every scan to initialize the SiV into the negatively charged state.

Fig.~\ref{dit}B shows the transmission and PLE spectra for a particular drop device in near resonance with two SiV centers, hereafter referred to as SiV 1 and 2.
All subsequent measurements are from this device --- data from other devices is included in Supplemental 3F.
The cavity is over coupled and has a total loss rate $\kappa/2\pi = 114$\,GHz corresponding to a total quality factor $Q=3540$ (Supplemental 3D).
The enhanced coupling of SiV 1 and 2 produces increased PSB fluorescence, which appears as sharp individual peaks within the comparatively dimmer inhomogeneous ensemble.
At the same time, the coherent coupling of the ZPL emission interferes with the transmitted excitation, resulting in a deep and narrow transmission dip at the SiV resonance frequency, which is a manifestation of dipole-induced transparency (DIT)~\cite{waks2006dit}.

DIT can also be utilized to probe the jumps in the SiV center charge state as shown in Fig.~\ref{dit}C.
Here, the $C$ transition of SiV 1 is resonantly driven and the transmission through the cavity is continuously monitored over many minutes.
The DIT dip vanishes when the SiV is ionized into a dark charge state, yielding an increased transmission rate.
Additional continuous excitation by a strongly attenuated 532-nm repump laser enables periodic re-ionization of the SiV back into the negatively charged state, reducing the transmission again via DIT.
The (re-)ionization rates can be directly tuned by the laser powers.
Thus, the bright and dark charge states can be reliably distinguished via single-shot scattering from the cavity (Fig.~\ref{dit}D) which is a precursor to the spin-dependent transmission discussed in the next section.

\begin{figure}[t]
    \centering
    \includegraphics{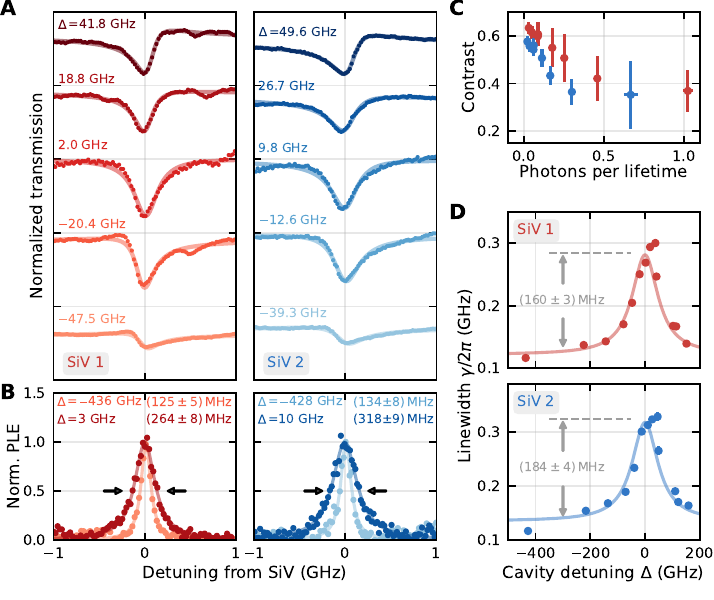}
    \caption{
        \textbf{High-cooperativity coupling.}
        (\textbf{A}) High-resolution transmission spectra at the SiV 1 and SiV 2 resonances for different SiV-cavity detunings.
        Fitting to the model enables determination of the cooperativity.
        (\textbf{B}) High-resolution PLE scans on SiV 1 and 2 showing broadened linewidths when the cavity is near resonant.
        (\textbf{C}) Dependence of the DIT transmission dip contrast near resonance as a function of input power.
        (\textbf{D}) Linewidths of SiV 1 and 2 as the cavity is tuned into and out of resonance.
        The fit assumes the measured cavity linewidth and is in good agreement with the data.
    }
    \label{high-cooperativity}
\end{figure}

To quantify the SiV-cavity coupling we fit the normalized DIT signal spectrum as a function of the detuning $\Delta = \omega_\mathrm{SiV} - \omega_\mathrm{cav}$ (Fig.~\ref{high-cooperativity}A).
The system is modeled via the input-output formalism~\cite{gardiner1985heisenberglangevin} with details provided in Supplemental 2A.
We note that thermalization between the ground states $\ket{1}$ and $\ket{2}$ (Fig.~\ref{dit}A) is fast at 4\,K compared to the relevant experimental time scales~\cite{jahnke2015sivelectronphonon} and so any optical pumping between the orbitals can be neglected.
However, the measured transmission spectrum will consequently be a classical average of the idealized two-level DIT spectra when the SiV is in $\ket{1}$ and $\ket{2}$ respectively, with the weights being given by the thermal populations $n_2/n_1 = \exp(-\Delta E/ k_B T)$.
At 4\,K and for a ground state splitting of approximately 50\,GHz (Fig.~\ref{dit}B), we have $n_2/n_1 \approx 0.54$ which significantly reduces the DIT dip contrast, giving the impression of a reduced vacuum Rabi frequency $g$.
Without accounting for this, fits of the DIT spectra to the two-level model (Supplemental 2A) then yield the intrinsic excited state loss rate $\gamma$ and a lower bound on $g$.
When SiV 1 is near resonance ($\Delta/2\pi = 0.523$\,GHz), we find $\gamma/2\pi = (110\pm1)$\,MHz and $g/2\pi > (2.13\pm0.01)$\,GHz, corresponding to a lower bound on the cooperativity $C = 4g^2/\kappa\gamma > 1.37\pm0.01$.
Likewise for SiV 2 ($\Delta/2\pi = 8.367$\,GHz) we find $\gamma/2\pi = (190\pm1)$\,MHz and $g/2\pi > (2.20\pm0.01)$\,GHz corresponding to $C > 0.86\pm0.01$.

We also measure the Purcell-enhanced broadening of the PLE linewidth as a function of detuning.
The linewidth broadening scales in proportion to the cavity lineshape and is maximized at $\gamma_\mathrm{cav} = 4g^2/\kappa$ when the cavity is on resonance (Supplemental 2A).
This constitutes an independent measurement of the Rabi frequency $g$ provided additional broadening mechanisms due to power or spectral diffusion are not present.
Fig.~\ref{high-cooperativity}B shows the PLE spectra of SiV 1 and 2 when the cavity is far detuned ($|\Delta|\gg \kappa$) and near resonance ($|\Delta|\ll \kappa$).
Fig.~\ref{high-cooperativity}D shows the linewidth as a function of the detuning $\Delta$.
Fitting the linewidth to the cavity lineshape enables determination of the peak broadening $\gamma_\mathrm{cav}$ and subsequently the cooperativity $C= \gamma_\mathrm{cav}/\gamma$.
For SiV 1 we find $\gamma_\mathrm{cav} = (160\pm3)$\,MHz corresponding to $C=1.32\pm0.04$.
Likewise for SiV 2, we find $\gamma_\mathrm{cav}/2\pi = (184\pm4)$\,MHz corresponding to $C = 1.36\pm0.4$.
These values are consistent with the DIT-based cooperativity bounds.

\section{Efficient spin-photon interactions}
\begin{figure}[t]
    \centering
    \includegraphics[width=\textwidth]{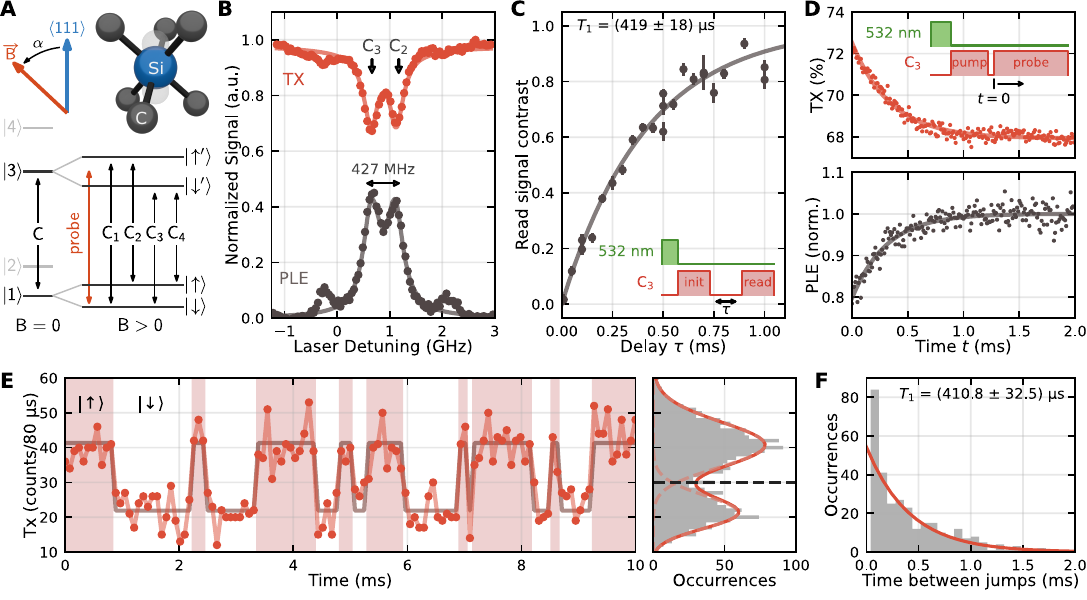}
    \caption{
        \textbf{An efficient spin-photon interface.}
        (\textbf{A}) States and optical transitions of an SiV center in an applied magnetic field at angle $\alpha$ from the symmetry axis.
        (\textbf{B}) Transmission and PLE spectra on SiV 1 with clear resolution of the spin conserving transitions $C_2$ and $C_3$.
        The two weaks peaks on either end are attributed to additional SiV centers.
        (\textbf{C}) Spin-relaxation curve obtained via PLE measurements on SiV 1, pulse sequence shown in the inset.
        (\textbf{D}) Spin-dependent switching of the cavity transmission and PLE signal.
        (\textbf{E}) A typical time trace of the cavity transmission demonstrating single shot readout of the quantum jumps of the SiV spin state (left).
        The intensity distribution of 15 time traces with a bimodal Poisson fit (right).
        The dashed line at 30\,cts/bin corresponds to the optimal discrimination threshold with fidelity $F=96$\%.
        (\textbf{F}) Distribution of the time between quantum jumps fit to an exponential distribution.
    }
    \label{spin-photon-interface}
\end{figure}

The sample is then placed on a custom mount containing a strong samarium-cobalt magnet which creates a static magnetic field of around 0.27\,T roughly aligned to the SiV symmetry axis at the device of interest (Supplemental 3G).
Under moderate strain, the effective gyromagnetic ratios of the ground and excited state manifolds differ~\cite{nguyen2019sivregister}, enabling resolution of the four distinct spin transitions (Fig.~\ref{spin-photon-interface}A,B).
When the angle $\alpha$ between the SiV axis and the applied field approaches zero, the spin-flipping optical transitions $C_1$ and $C_4$ become suppressed while the spin-conserving transitions $C_2$ and $C_3$ become highly cyclic~\cite{sukachev2017siv}.
Simultaneously, the spin relaxation time $T_1$ rapidly increases, reaching up to a few milliseconds at 4\,K~\cite{rogers2014sivspin}.
We probe the spin $T_1$ by optically pumping the $C_3$ transition of the SiV (initializing into $\ket{\uparrow}$) then allowing it to thermally relax in the dark for a time $\tau$ before probing the $\ket{\downarrow}$ population with a second read pulse as shown in Fig.~\ref{spin-photon-interface}C.
Imperfect alignment of the sample during mounting leads to a simulated misalignment angle of $\alpha \approx 4^\circ$, with initial spin relaxation times of $T_1 = (97\pm6)$\,\textmu s and $(38\pm3)$\,\textmu s for SiV 1 and 2 respectively.
We utilize an neodymium magnet outside of the cryostat to further optimize the field alignment resulting in a final simulated misalignment angle $\alpha < 1^\circ$ and measured $T_1 = (419\pm18)$\,\textmu s for SiV 1 (Supplemental 3H).
SiV 2 was not measured further due to its reduced $T_1$ and DIT contrast relative to SiV 1.

We then perform pump-probe measurements on SiV 1 to analyze the modulation of the cavity transmission due to the spin state.
The resonant laser (tuned to the $C_3$ transition) is split into two paths with separate pulse control and attenuation before being recombined (Supplemental 3A).
This enables efficient optical pumping during the pump step and ensures that the probe is sufficiently weak as to not affect the spin dynamics.
In Fig.~\ref{spin-photon-interface}D, we plot the average transmission and PLE signal from SiV 1 as the spin thermally equilibrates.
During the pump step, SiV 1 is intialized into $\ket{\uparrow}$ causing an average reduction in the DIT contrast and PLE signal.
Both the DIT contrast and PLE signal increases as the spin relaxes into thermal equilibrium.
Fitting the recovery rate yields a characteristic time of $(310\pm20)$\,\textmu s and $(349\pm9)$\,\textmu s for the transmission and PLE curves respectively, which are both roughly consistent with the prior measurement of $T_1$.
The measured spin initialization fidelity is determined to be $(20\pm1)$\%, which we hypothesize is limited by the green-repump-induced spectral diffusion (Supplemental 3I).

Finally, we utilize the same pump-probe sequence to observe the quantum jumps of the SiV spin state via single-shot readout as shown in Fig.~\ref{spin-photon-interface}E.
We then select 15 separate measurement sequences and fit the intensity distribution to a bimodal Poisson distribution.
The optimal discrimination threshold is evaluated to be 30\,counts/shot which yields a single-shot readout fidelity of $F = (96.0\pm0.3)$\% (Supplemental 3I), limited by the microscope efficiency (Supplemental 3E).
The time between subsequent jumps are binned into a histogram shown in Fig.~\ref{spin-photon-interface}F.
We fit the distribution neglecting the first bin (which captures much of the ``false jumps'' arising from signal noise near the threshold) to estimate of the relaxation time at $T_1 = (411\pm33)$\,\textmu s in good agreement with the our earlier measurements.

\section{Outlook}
The devices presented here are the first demonstration of high-cooperativity ($C>1$) coupling to quantum emitters in a planar, hybrid nanophotonic platform.
These devices already have cooperativities on par with those used in other cQED-based networking experiments utilizing monolithic cavities~\cite{stas2022sivregister,bersin2024sivnetworking,knaut2024sivnetwork}, indicating the potential for near-term quantum networking with GaP-on-diamond photonics.
Moreover, these devices are far from theoretical limits indicating significant potential for improvement.
Increased robustness to fabrication imperfections by refinement of the fabrication techniques and design~\cite{abulnaga2025robustphc} could yield a 10-fold increase in measured quality factors.
Such increases are still well within the limits of hybrid-integrated GaP devices for which measured quality factors exceed $10^5$~\cite{yama2024bgap}.
Alternatively, techniques for near-surface defect formation, e.g.\ \textit{in situ} doping~\cite{ohno2012deltadoping}, could enable defect formation as close as 5\,nm beneath the diamond surface, yielding an approximately 50\% larger nominal overlap factor.
The subsequent relaxation of the overlap constraint would then enable larger $Q/V$ designs, increasing the theoretical limit for optimized performance.

Our work demonstrates that GaP-on-diamond photonics is an ideal platform for the realization of scalable quantum networking technologies.
The strong nonlinear properties of GaP can be utilized for on-chip quantum frequency conversion~\cite{logan2023triply}, modified protocols utilizing squeezed light~\cite{le2025squeezed}, and simultaneous integration of phononics for spin control~\cite{lemonde2018sivphonon} and cavity-optomechanics~\cite{xinyuan2023gapoptomech}.
Importantly, the lack of diamond etching and compatibility with surface treatments~\cite{pederson2024surface} could enable the use of alternative defects with superior properties --- including those without inversion symmetry such as the NV center.
Altogether, these results establish GaP-on-diamond as a promising platform for the development of highly scalable, next-generation quantum technologies.

\section*{Acknowledgments}
The authors thank N.~Shane Patrick, Doc Daugherty, Mark Morgan, Mark Brunson, and Grace Tong for assistance with lithography and etching.
The authors also thank Essance Ray and Xiaodong Xu for training and provision of the stamp transfer stage.
This material is primarily based upon work supported by Department of Energy, Office of Science, National Quantum Information Science Research Centers, Co-design Center for Quantum Advantage (C2QA) under contract number DE-SC0012704 (Pacific Northwest National Laboratory, FWP 76274).
N.S.Y.\ was supported by the National Science Foundation Graduate Research Fellowship Program under Grant No.~DGE-2140004.
Part of this work was conducted at the Washington Nanofabrication Facility/Molecular Analysis Facility, a National Nanotechnology Coordinated Infrastructure (NNCI) site at the University of Washington with partial support from the National Science Foundation via awards NNCI-1542101 and NNCI-2025489.

\bibliography{main}

\begin{thebibliography}{10}

\bibitem{stas2022sivregister}
P.-J. Stas, Y.~Q. Huan, B.~Machielse, E.~N. Knall, A.~Suleymanzade, B.~Pingault, M.~Sutula, S.~W. Ding, C.~M. Knaut, D.~R. Assumpcao, Y.-C. Wei, M.~K. Bhaskar, R.~Riedinger, D.~D. Sukachev, H.~Park, M.~Lončar, D.~S. Levonian, and M.~D. Lukin, ``Robust multi-qubit quantum network node with integrated error detection,'' {\em Science}, vol.~378, no.~6619, pp.~557--560, 2022.

\bibitem{bersin2024sivnetworking}
E.~Bersin, M.~Sutula, Y.~Q. Huan, A.~Suleymanzade, D.~R. Assumpcao, Y.-C. Wei, P.-J. Stas, C.~M. Knaut, E.~N. Knall, C.~Langrock, N.~Sinclair, R.~Murphy, R.~Riedinger, M.~Yeh, C.~Xin, S.~Bandyopadhyay, D.~D. Sukachev, B.~Machielse, D.~S. Levonian, M.~K. Bhaskar, S.~Hamilton, H.~Park, M.~Lon\ifmmode~\check{c}\else \v{c}\fi{}ar, M.~M. Fejer, P.~B. Dixon, D.~R. Englund, and M.~D. Lukin, ``Telecom networking with a diamond quantum memory,'' {\em PRX Quantum}, vol.~5, p.~010303, Jan 2024.

\bibitem{knaut2024sivnetwork}
C.~M. Knaut, A.~Suleymanzade, Y.-C. Wei, D.~R. Assumpcao, P.-J. Stas, Y.~Q. Huan, B.~Machielse, E.~N. Knall, M.~Sutula, G.~Baranes, N.~Sinclair, C.~De-Eknamkul, D.~S. Levonian, M.~K. Bhaskar, H.~Park, M.~Lončar, and M.~D. Lukin, ``Entanglement of nanophotonic quantum memory nodes in a telecom network,'' {\em Nature}, vol.~629, p.~573–578, May 2024.

\bibitem{burek2014diamondcavities}
M.~J. Burek, Y.~Chu, M.~S.~Z. Liddy, P.~Patel, J.~Rochman, S.~Meesala, W.~Hong, Q.~Quan, M.~D. Lukin, and M.~Lon{\v{c}}ar, ``High quality-factor optical nanocavities in bulk single-crystal diamond,'' {\em Nature Communications}, vol.~5, p.~5718, Dec 2014.

\bibitem{sipahigil2016phc}
A.~Sipahigil, R.~E. Evans, D.~D. Sukachev, M.~J. Burek, J.~Borregaard, M.~K. Bhaskar, C.~T. Nguyen, J.~L. Pacheco, H.~A. Atikian, C.~Meuwly, R.~M. Camacho, F.~Jelezko, E.~Bielejec, H.~Park, M.~Lončar, and M.~D. Lukin, ``An integrated diamond nanophotonics platform for quantum-optical networks,'' {\em Science}, vol.~354, no.~6314, pp.~847--850, 2016.

\bibitem{nguyen2019sivregister}
C.~T. Nguyen, D.~D. Sukachev, M.~K. Bhaskar, B.~Machielse, D.~S. Levonian, E.~N. Knall, P.~Stroganov, C.~Chia, M.~J. Burek, R.~Riedinger, H.~Park, M.~Lon\ifmmode~\check{c}\else \v{c}\fi{}ar, and M.~D. Lukin, ``An integrated nanophotonic quantum register based on silicon-vacancy spins in diamond,'' {\em Phys. Rev. B}, vol.~100, p.~165428, Oct 2019.

\bibitem{ding2024diamondmembrane}
S.~W. Ding, M.~Haas, X.~Guo, K.~Kuruma, C.~Jin, Z.~Li, D.~D. Awschalom, N.~Delegan, F.~J. Heremans, A.~A. High, and M.~Loncar, ``High-q cavity interface for color centers in thin film diamond,'' {\em Nature Communications}, vol.~15, p.~6358, Jul 2024.

\bibitem{riedel2025lightsynq}
D.~Riedel, T.~Graziosi, Z.~Wang, C.~De-Eknamkul, A.~Abulnaga, J.~Dietz, A.~Mucchietto, M.~Haas, M.~Sutula, P.~Barral, M.~Pompili, M.~Raha, C.~Robens, J.~Ha, D.~Sukachev, D.~Levonian, M.~Bhaskar, M.~Markham, and B.~Machielse, ``A scalable photonic quantum interconnect platform,'' 2025.

\bibitem{cui2015plasmadamage}
S.~Cui, A.~S. Greenspon, K.~Ohno, B.~A. Myers, A.~C.~B. Jayich, D.~D. Awschalom, and E.~L. Hu, ``Reduced plasma-induced damage to near-surface nitrogen-vacancy centers in diamond,'' {\em Nano Letters}, vol.~15, pp.~2887--2891, 05 2015.

\bibitem{chakravarthi2021nvsurface}
S.~Chakravarthi, C.~Pederson, Z.~Kazi, A.~Ivanov, and K.-M.~C. Fu, ``Impact of surface and laser-induced noise on the spectral stability of implanted nitrogen-vacancy centers in diamond,'' {\em Physical Review B}, vol.~104, no.~8, p.~085425, 2021.

\bibitem{sipahigil2014indistinguishable}
A.~Sipahigil, K.~D. Jahnke, L.~J. Rogers, T.~Teraji, J.~Isoya, A.~S. Zibrov, F.~Jelezko, and M.~D. Lukin, ``Indistinguishable {Photons} from {Separated} {Silicon}-{Vacancy} {Centers} in {Diamond},'' {\em Physical Review Letters}, vol.~113, p.~113602, Sept. 2014.

\bibitem{wan2020large}
N.~H. Wan, T.-J. Lu, K.~C. Chen, M.~P. Walsh, M.~E. Trusheim, L.~De~Santis, E.~A. Bersin, I.~B. Harris, S.~L. Mouradian, I.~R. Christen, {\em et~al.}, ``Large-scale integration of artificial atoms in hybrid photonic circuits,'' {\em Nature}, vol.~583, no.~7815, pp.~226--231, 2020.

\bibitem{riedel2023linbo3}
D.~Riedel, H.~Lee, J.~F. Herrmann, J.~Grzesik, V.~Ansari, J.-M. Borit, H.~S. Stokowski, S.~Aghaeimeibodi, H.~Lu, P.~J. McQuade, N.~A. Melosh, Z.-X. Shen, A.~H. Safavi-Naeini, and J.~Vu{\v{c}}kovi{\'{c}}, ``Efficient photonic integration of diamond color centers and thin-film lithium niobate,'' {\em ACS Photonics}, vol.~10, pp.~4236--4243, Dec 2023.

\bibitem{li2024cmos}
L.~Li, L.~D. Santis, I.~B.~W. Harris, K.~C. Chen, Y.~Gao, I.~Christen, H.~Choi, M.~Trusheim, Y.~Song, C.~Errando-Herranz, J.~Du, Y.~Hu, G.~Clark, M.~I. Ibrahim, G.~Gilbert, R.~Han, and D.~Englund, ``Heterogeneous integration of spin--photon interfaces with a cmos platform,'' {\em Nature}, vol.~630, pp.~70--76, Jun 2024.

\bibitem{barclay2009gapphc}
P.~E. Barclay, K.-M. Fu, C.~Santori, and R.~G. Beausoleil, ``Hybrid photonic crystal cavity and waveguide for coupling to diamond nv-centers,'' {\em Opt. Express}, vol.~17, pp.~9588--9601, Jun 2009.

\bibitem{gould2015gap}
M.~Gould, S.~Chakravarthi, I.~R. Christen, N.~Thomas, S.~Dadgostar, Y.~Song, M.~L. Lee, F.~Hatami, and K.-M.~C. Fu, ``Large-scale gap-on-diamond integrated photonics platform for nv center-based quantum information,'' {\em J. Opt. Soc. Am. B}, vol.~33, pp.~B35--B42, Mar 2016.

\bibitem{chakravarthi2023stamp}
S.~Chakravarthi, N.~S. Yama, A.~Abulnaga, D.~Huang, C.~Pederson, K.~Hestroffer, F.~Hatami, N.~P. de~Leon, and K.-M.~C. Fu, ``Hybrid integration of gap photonic crystal cavities with silicon-vacancy centers in diamond by stamp-transfer,'' {\em Nano Letters}, vol.~23, pp.~3708--3715, 05 2023.

\bibitem{corso1996gapChi2}
A.~D. Corso, F.~Mauri, and A.~Rubio, ``Density-functional theory of the nonlinear optical susceptibility: Application to cubic semiconductors,'' {\em Phys. Rev. B}, vol.~53, pp.~15638--15642, Jun 1996.

\bibitem{wilson2020nonlineargap}
D.~J. Wilson, K.~Schneider, S.~H{\"o}nl, M.~Anderson, Y.~Baumgartner, L.~Czornomaz, T.~J. Kippenberg, and P.~Seidler, ``Integrated gallium phosphide nonlinear photonics,'' {\em Nature Photonics}, vol.~14, no.~1, pp.~57--62, 2020.

\bibitem{honl2022gaptransduction}
S.~H{\"o}nl, Y.~Popoff, D.~Caimi, A.~Beccari, T.~J. Kippenberg, and P.~Seidler, ``Microwave-to-optical conversion with a gallium phosphide photonic crystal cavity,'' {\em Nature Communications}, vol.~13, no.~1, p.~2065, 2022.

\bibitem{logan2023triply}
A.~D. Logan, S.~Shree, S.~Chakravarthi, N.~Yama, C.~Pederson, K.~Hestroffer, F.~Hatami, and K.-M.~C. Fu, ``Triply-resonant sum frequency conversion with gallium phosphide ring resonators,'' {\em Opt. Express}, vol.~31, pp.~1516--1531, Jan 2023.

\bibitem{mytsyk2015gapphotoelastic}
B.~G. Mytsyk, N.~M. Demyanyshyn, and O.~M. Sakharuk, ``Elasto-optic effect anisotropy in gallium phosphide crystals,'' {\em Appl. Opt.}, vol.~54, pp.~8546--8553, Oct 2015.

\bibitem{schneider2019gapoptomech}
K.~Schneider, Y.~Baumgartner, S.~H\"{o}nl, P.~Welter, H.~Hahn, D.~J. Wilson, L.~Czornomaz, and P.~Seidler, ``Optomechanics with one-dimensional gallium phosphide photonic crystal cavities,'' {\em Optica}, vol.~6, pp.~577--584, May 2019.

\bibitem{xinyuan2023gapoptomech}
X.~Ma, P.~K. Shandilya, and P.~E. Barclay, ``Semiconductor-on-diamond cavities for spin optomechanics,'' {\em Opt. Express}, vol.~31, pp.~22470--22480, Jul 2023.

\bibitem{yama2024bgap}
N.~S. Yama, I.-T. Chen, S.~Chakravarthi, B.~Li, C.~Pederson, B.~E. Matthews, S.~R. Spurgeon, D.~E. Perea, M.~G. Wirth, P.~V. Sushko, M.~Li, and K.-M.~C. Fu, ``Silicon-lattice-matched boron-doped gallium phosphide: A scalable acousto-optic platform,'' {\em Advanced Materials}, vol.~36, no.~5, p.~2305434, 2024.

\bibitem{dibos2018hybrid}
A.~M. Dibos, M.~Raha, C.~M. Phenicie, and J.~D. Thompson, ``Atomic source of single photons in the telecom band,'' {\em Phys. Rev. Lett.}, vol.~120, p.~243601, Jun 2018.

\bibitem{ourari2023hybrid}
S.~Ourari, {\L}.~Dusanowski, S.~P. Horvath, M.~T. Uysal, C.~M. Phenicie, P.~Stevenson, M.~Raha, S.~Chen, R.~J. Cava, N.~P. de~Leon, and J.~D. Thompson, ``Indistinguishable telecom band photons from a single er ion in the solid state,'' {\em Nature}, vol.~620, pp.~977--981, Aug 2023.

\bibitem{wu2023hybrid}
C.-J. Wu, D.~Riedel, A.~Ruskuc, D.~Zhong, H.~Kwon, and A.~Faraon, ``Near-infrared hybrid quantum photonic interface for ${}^{171}{\mathrm{yb}}^{3+}$ solid-state qubits,'' {\em Phys. Rev. Appl.}, vol.~20, p.~044018, Oct 2023.

\bibitem{quan2010phcdesign}
Q.~Quan, P.~B. Deotare, and M.~Loncar, ``{Photonic crystal nanobeam cavity strongly coupled to the feeding waveguide},'' {\em Applied Physics Letters}, vol.~96, p.~203102, 05 2010.

\bibitem{quan2011phcdesign}
Q.~Quan and M.~Loncar, ``Deterministic design of wavelength scale, ultra-high q photonic crystal nanobeam cavities,'' {\em Opt. Express}, vol.~19, pp.~18529--18542, Sep 2011.

\bibitem{waks2006dit}
E.~Waks and J.~Vuckovic, ``Dipole induced transparency in drop-filter cavity-waveguide systems,'' {\em Phys. Rev. Lett.}, vol.~96, p.~153601, Apr 2006.

\bibitem{gardiner1985heisenberglangevin}
C.~W. Gardiner and M.~J. Collett, ``Input and output in damped quantum systems: Quantum stochastic differential equations and the master equation,'' {\em Phys. Rev. A}, vol.~31, pp.~3761--3774, Jun 1985.

\bibitem{jahnke2015sivelectronphonon}
K.~D. Jahnke, A.~Sipahigil, J.~M. Binder, M.~W. Doherty, M.~Metsch, L.~J. Rogers, N.~B. Manson, M.~D. Lukin, and F.~Jelezko, ``Electron--phonon processes of the silicon-vacancy centre in diamond,'' {\em New Journal of Physics}, vol.~17, no.~4, p.~043011, 2015.

\bibitem{sukachev2017siv}
D.~D. Sukachev, A.~Sipahigil, C.~T. Nguyen, M.~K. Bhaskar, R.~E. Evans, F.~Jelezko, and M.~D. Lukin, ``Silicon-vacancy spin qubit in diamond: A quantum memory exceeding 10 ms with single-shot state readout,'' {\em Phys. Rev. Lett.}, vol.~119, p.~223602, Nov 2017.

\bibitem{rogers2014sivspin}
L.~J. Rogers, K.~D. Jahnke, M.~H. Metsch, A.~Sipahigil, J.~M. Binder, T.~Teraji, H.~Sumiya, J.~Isoya, M.~D. Lukin, P.~Hemmer, and F.~Jelezko, ``All-optical initialization, readout, and coherent preparation of single silicon-vacancy spins in diamond,'' {\em Phys. Rev. Lett.}, vol.~113, p.~263602, Dec 2014.

\bibitem{abulnaga2025robustphc}
A.~Abulnaga, S.~Karg, S.~Mukherjee, A.~Gupta, K.~W. Baldwin, L.~N. Pfeiffer, and N.~P. de~Leon, ``Design and fabrication of robust hybrid photonic crystal cavities,'' {\em Nanophotonics}, vol.~14, no.~11, pp.~1927--1937, 2025.

\bibitem{ohno2012deltadoping}
K.~Ohno, F.~Joseph~Heremans, L.~C. Bassett, B.~A. Myers, D.~M. Toyli, A.~C. Bleszynski~Jayich, C.~J. Palmstrøm, and D.~D. Awschalom, ``Engineering shallow spins in diamond with nitrogen delta-doping,'' {\em Applied Physics Letters}, vol.~101, p.~082413, 08 2012.

\bibitem{le2025squeezed}
T.~K. L\^e, D.~M. Lukin, C.~Roques-Carmes, A.~Karnieli, E.~Lustig, M.~A. Guidry, S.~Fan, and J.~Vu\ifmmode \check{c}\else \v{c}\fi{}kovi\ifmmode~\acute{c}\else \'{c}\fi{}, ``Cavity quantum electrodynamics in a finite-bandwidth squeezed reservoir,'' {\em Phys. Rev. Appl.}, vol.~24, p.~034053, Sep 2025.

\bibitem{lemonde2018sivphonon}
M.-A. Lemonde, S.~Meesala, A.~Sipahigil, M.~J.~A. Schuetz, M.~D. Lukin, M.~Loncar, and P.~Rabl, ``Phonon networks with silicon-vacancy centers in diamond waveguides,'' {\em Phys. Rev. Lett.}, vol.~120, p.~213603, May 2018.

\bibitem{pederson2024surface}
C.~Pederson, R.~Giridharagopal, F.~Zhao, S.~T. Dunham, Y.~Raitses, D.~S. Ginger, and K.-M.~C. Fu, ``Optical tuning of the diamond fermi level measured by correlated scanning probe microscopy and quantum defect spectroscopy,'' {\em Phys. Rev. Mater.}, vol.~8, p.~036201, Mar 2024.

\end{thebibliography}


\begin{thebibliography}{10}

\bibitem{quan2010phcdesign}
Q.~Quan, P.~B. Deotare, and M.~Loncar, ``{Photonic crystal nanobeam cavity strongly coupled to the feeding waveguide},'' {\em Applied Physics Letters}, vol.~96, p.~203102, 05 2010.

\bibitem{quan2011phcdesign}
Q.~Quan and M.~Loncar, ``Deterministic design of wavelength scale, ultra-high q photonic crystal nanobeam cavities,'' {\em Opt. Express}, vol.~19, pp.~18529--18542, Sep 2011.

\bibitem{johnson2001mpb}
S.~G. Johnson and J.~D. Joannopoulos, ``Block-iterative frequency-domain methods for maxwell's equations in a planewave basis,'' {\em Opt. Express}, vol.~8, pp.~173--190, Jan 2001.

\bibitem{malherbe2017lipo}
C.~Malherbe and N.~Vayatis, ``Global optimization of lipschitz functions,'' 2017.

\bibitem{king2009dlib}
D.~E. King, ``Dlib-ml: A machine learning toolkit,'' {\em Journal of Machine Learning Research}, vol.~10, pp.~1755--1758, 2009.

\bibitem{chakravarthi2023stamp}
S.~Chakravarthi, N.~S. Yama, A.~Abulnaga, D.~Huang, C.~Pederson, K.~Hestroffer, F.~Hatami, N.~P. de~Leon, and K.-M.~C. Fu, ``Hybrid integration of gap photonic crystal cavities with silicon-vacancy centers in diamond by stamp-transfer,'' {\em Nano Letters}, vol.~23, pp.~3708--3715, 05 2023.

\bibitem{gardiner1985heisenberglangevin}
C.~W. Gardiner and M.~J. Collett, ``Input and output in damped quantum systems: Quantum stochastic differential equations and the master equation,'' {\em Phys. Rev. A}, vol.~31, pp.~3761--3774, Jun 1985.

\bibitem{kubo1962cumulant}
R.~Kubo, ``Generalized cumulant expansion method,'' {\em Journal of the Physical Society of Japan}, vol.~17, no.~7, pp.~1100--1120, 1962.

\bibitem{sanchez2020cumulant}
M.~Sánchez-Barquilla, R.~E.~F. Silva, and J.~Feist, ``Cumulant expansion for the treatment of light–matter interactions in arbitrary material structures,'' {\em The Journal of Chemical Physics}, vol.~152, p.~034108, 01 2020.

\bibitem{waks2006dit}
E.~Waks and J.~Vuckovic, ``Dipole induced transparency in drop-filter cavity-waveguide systems,'' {\em Phys. Rev. Lett.}, vol.~96, p.~153601, Apr 2006.

\bibitem{jahnke2015sivelectronphonon}
K.~D. Jahnke, A.~Sipahigil, J.~M. Binder, M.~W. Doherty, M.~Metsch, L.~J. Rogers, N.~B. Manson, M.~D. Lukin, and F.~Jelezko, ``Electron--phonon processes of the silicon-vacancy centre in diamond,'' {\em New Journal of Physics}, vol.~17, no.~4, p.~043011, 2015.

\bibitem{ortner2020magpylib}
M.~Ortner and L.~G.~C. Bandeira, ``Magpylib: A free python package for magnetic field computation,'' {\em SoftwareX}, vol.~11, p.~100466, 2020.

\bibitem{robledo2011fidelity}
L.~Robledo, L.~Childress, H.~Bernien, B.~Hensen, P.~F.~A. Alkemade, and R.~Hanson, ``High-fidelity projective read-out of a solid-state spin quantum register,'' {\em Nature}, vol.~477, pp.~574--578, Sep 2011.

\end{thebibliography}

\end{document}


\title{Supplementary information for ``A scalable gallium-phosphide-on-diamond spin-photon interface''}

\author{Nicholas S. Yama}
\affiliation{University of Washington, Electrical and Computer Engineering Department, Seattle, WA, 98105, USA}
\author{Chun-Chi Wu}
\affiliation{University of Washington, Electrical and Computer Engineering Department, Seattle, WA, 98105, USA}
\author{Fariba Hatami}
\affiliation{Department of Physics, Humboldt-Universitat zu Berlin, Newtonstrasse, Berlin, 10117, Germany}
\author{Kai-Mei C. Fu}
\affiliation{University of Washington, Electrical and Computer Engineering Department, Seattle, WA, 98105, USA}
\affiliation{University of Washington, Physics Department, Seattle, WA, 98105, USA}
\affiliation{Physical Sciences Division, Pacific Northwest National Laboratory, Richland, Washington 99352, USA}

\date{\today}

\maketitle            
\tableofcontents

\clearpage
\section{Nanophotonic cavity design, optimization, and fabrication}

\subsection{Nanophotonic design}
\begin{figure}[h]
    \centering
    \includegraphics{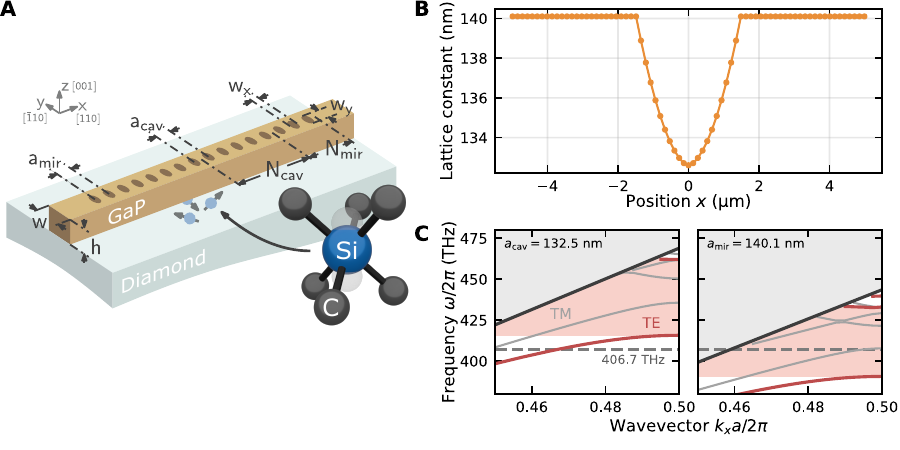}
    \caption{
        \textbf{Nanobeam cavity design.}
        (\textbf{A}) Illustration of the cavity design with labeled device parameters.
        (\textbf{B}) Lattice constant as a function of position for the optimized cavity design with $N_\mathrm{cav} = 12$ and $N_\mathrm{mir} = 25$.
        (\textbf{C}) Photonic band structures at the center (left) and in the mirror region (right) of the optimized cavity.
    }
    \label{fig:si:design}
\end{figure}

The cavity is formed from a 1-d gallium phosphide (GaP) photonic crystal (PhC) nanobeam cavity heterogeneously integrated on a planar diamond substrate as shown in Fig.~\ref{fig:si:design}.
The nanobeam cavity consists of a GaP waveguide of width $w$ and height $h$ with a series of identical elliptical holes symmetrically arranged along the waveguide length.
The holes have minor/major axes denoted by $w_x$ and $w_y$ respectively.
The lattice constant $a$ is quadratically chirped along the waveguide length to form a defect mode while sections of constant spacing are formed on either end which serve as mirrors.
Quadratic chirping is chosen as it is known to minimize the mode volume and radiation losses simultaneously~\cite{quan2010phcdesign,quan2011phcdesign}.
Specifically, the center-to-center spacing of the holes is varied quadratically as
\begin{equation*}
    a_n = a_{\mathrm{cav}} + \frac{(a_{\mathrm{mir}} - a_{\mathrm{cav}}) n^2}{(N_{\mathrm{cav}} - 1)^2},
\end{equation*}
where $n$ is the hole index ranging from $0$ to the number of holes $N_{\mathrm{cav}}-1$ where $N_{\mathrm{cav}}$ is the number of holes in the tapered section on either end.
The lattice constants $a_{\mathrm{mir}}$ and $a_{\mathrm{cav}}$ denote the mirror region and cavity-center lattice constants.
In addition to the taper, we include an additional $N_{\mathrm{mir}}$ holes on either end with constant spacing $a_{\mathrm{mir}}$.
Parameterizing the design in terms of the variables $(w,h,w_x,w_y,a_{\mathrm{cav}},a_{\mathrm{mir}},N_{\mathrm{cav}},N_{\mathrm{mir}})$ then enables numerical optimization.

Within certain domains of the parameter space, this structure supports a small number of discrete transverse-electric (TE) resonances within the photonic bandgap that can support large $Q/V$ ratios.
When the cavity is aligned along one of the $\langle110\rangle$ axes of the $(001)$-diamond, the SiV centers that are orthogonal to the cavity axis have a large transition dipole moment that is aligned to the TE field (with a $35^\circ$ offset due to the vertical component of the SiV dipole).
The structure also supports a lossy continuum of transverse-magnetic (TM) modes which only weakly couple to the SiV transitions.
For finite numbers of mirror holes $N_{\mathrm{cav}}$, photons in the cavity can couple to the continuum of waveguide modes on either end of the cavity to allow for input/output --- this is the ``mirror-coupled'' or ``drop'' design configuration.
We also fabricate devices in which such scattering is suppressed via increasing $N_{\mathrm{cav}}$ and light is instead coupled in/out of the cavity mode via an adjacent waveguide --- this is the ``side-coupled'' or ``thru'' design configuration.
In either case, the fundamental mode (generally the highest-frequency resonance) has a quasi-Gaussian field profile along the cavity length and possesses the largest $Q/V$ ratio.
Consequently it is the fundamental mode which is utilized for SiV coupling via the evanescent fields that penetrate into the diamond substrate.
In our case, the SiV centers are largely concentrated at 20\,nm beneath the interface, well within the $1/e$ field intensity decay length of roughly 40\,nm, allowing for strong coupling despite not being at the field maximum.

\clearpage
\subsection{Design optimization}
The primary design objective is to maximize the cooperativity
\begin{equation}
    \label{eq:cooperativity}
    C =
    \frac{3}{4\pi} 
    \bigg(\frac{\gamma_0}{\gamma}\bigg)
    \bigg(\frac{Q}{V}\bigg) 
    \bigg( \frac{\bm\mu}{|\bm\mu|} \cdot \frac{\vb E(\vb x_a)}{\max|\vb E|} \bigg)^2,
\end{equation}
where $\gamma_0/\gamma$ gives the intrinsic radiative efficiency of the target transition, $Q$ is the cavity mode quality factor and $V$ is the effective-wavelength-normalized mode volume,
\begin{equation*}
    V = \bigg(\frac{n_{\mathrm{GaP}}}{\lambda_0}\bigg)^3 \frac{\int\dd[3]{\vb x} n^2(\vb x) |\vb E(\vb x)|^2 }{\max(n^2(\vb x) |\vb E(\vb x)|^2)},
\end{equation*}
with $n(\vb x)$ and $\vb E(\vb x)$ as the refractive index and modal field respectively; $n_{\mathrm{GaP}}$ is the GaP refractive index and $\lambda_0$ is the resonant wavelength of the mode.
The last factor describes the overlap of the SiV transition dipole moment $\bm\mu$ (located at $\vb x_a$) and the cavity mode field profile $\vb E(\vb x)$.

In hybrid devices the overlap factor is itself an important figure of merit whose optimization must be balanced with $Q/V$.
For example, strict maximization of $Q/V$ will tend to yield mode profiles that are increasingly localized within the GaP waveguiding layer, ultimately diminishing the cooperativity.
At the same time, attempting to increase the evanescent field amplitude (and thus the overlap factor) will generally diminish $Q$ (via increased scattering into the substrate) and enlarge $V$ (due to the lower effective index of the mode).

Our design optimization consists of three stages: (1) a forward-design process in which $Q/V$ is theoretically optimized, (2) an inverse-design process where the parameter space is algorithmically searched to maximize the cooperativity $C$, and (3) manual optimization of the input/output coupling.
These steps are detailed below.

\subsubsection{Forward design}
The forward-design process utilizes well-established theoretical heuristics for photonic nanobeam cavity design~\cite{quan2010phcdesign,quan2011phcdesign}.
While this typically neglects the geometric factor in favor of simply maximizing $Q/V$, the significantly sub-wavelength depth of the SiV ensures that the geometric factor remains roughly within an order of magnitude for most design geometries of interest.
In contrast, $Q/V$ can scale exponentially making this a good first step.

We begin by simulating the photonic band structures of 1-d GaP-on-diamond photonic crystal (PhC) waveguide structures using MIT Photonic Bands (MPB)~\cite{johnson2001mpb}.
The GaP and diamond refractive indices are taken to be $n_\mathrm{GaP} = 3.22$ and $n_\mathrm{dia} = 2.406$ respectively.
We perform coarse sweeps over all unit cell parameters $(w,h,w_x,w_y)$, scaled by the lattice constant $a$, to determine the regions of the parameter space that give rise to large bandgap-midgap frequency ratios (large ``mirror strengths''~\cite{quan2011phcdesign}).
Within a region of interest, a fine sweep of the parameters is formed to optimize the bandgap-midgap ratio from which a handful of candidate ``cavity'' unit cell designs are chosen.
Following the standard approach, the expected cavity resonance frequency ($737$\,nm or $406.774$\,THz) should lie slightly below the bottom of the bandgap which allows us to fix the cavity lattice constant $a_{\mathrm{cav}}$ and the other unit-cell parameters to a specific length in nanometers.
We then perform a sweep of potential mirror lattice constant $a_{\mathrm{mir}}$ while maintaining the dimensions $(w,h,w_x,w_y)$.
An increase in the lattice constant pulls down the bands towards lower absolute frequencies until the target frequency lies at the middle of the bandgap.
This maximizes the effective mirror strength at the target frequency and so the corresponding lattice constant is chosen as $a_{\mathrm{mir}}$.
Each set of parameters $(w,h,w_x,w_y,a_{\mathrm{cav}},a_{\mathrm{mir}})$ then constitutes to a particular candidate design.

We then simulate each of the candidate designs in finite-difference-time-domain (Lumerical FDTD) with varying numbers of cavity $N_{\mathrm{cav}}$ and mirror $N_{\mathrm{mir}}$ holes to verify the expected theoretical results.
We find that $N_{\mathrm{cav}}=12$ provides a good balance of increasing the $Q$ factor without also increasing the mode volume significantly.
As such, all devices hereafter are designed with $N_{\mathrm{cav}}=12$.
On the other hand, $N_{\mathrm{mir}}$ can vary to modulate the waveguide coupling.
We find that $Q$ factor begins to saturate at around $N_{\mathrm{mir}}=10$ indicating that the coupling loss rate is approaching the intrinsic loss rate.
To compare designs we consider only the intrinsic loss and so we temporarily set $N_{\mathrm{mir}}=25$ to eliminate the waveguide coupling.

\subsubsection{Inverse design}
We then implement an inverse design parameter search to further optimize the device cooperativity $C$.
The search aims to optimize the objective function $\eta$ over the parameter space spanned by $(w,h,w_x,w_y,a_{\mathrm{cav}},a_{\mathrm{mir}})$, with
\begin{equation}
    \eta = \frac{Q_\mathrm{fab} Q_\mathrm{sim}}{V(Q_\mathrm{fab} + Q_\mathrm{sim})} 
    \abs{ \frac{E_y(\vb x_a)}{\max|\vb E|} },
\end{equation}
which is proportional to the cooperativity Eq.~\eqref{eq:cooperativity} with $\bm\mu/|\bm\mu| = \vu y$ and $1/Q = (1/Q_\mathrm{fab}) + (1/Q_\mathrm{sim})$.
The artificial fabrication-limited quality factor $Q_\mathrm{fab} = 5\times10^4$ is imposed to bias the search against unrealistic radiation quality factors $Q_\mathrm{sim} > 10^5$ which would skew the results.
This limit is of course somewhat arbitrary and, with improved fabrication methods, one may be able to relax this constraint.
Because even slight changes in the cavity parameters can drastically design shift the resonance wavelength $\lambda_0$, the evaluation of the field component $E_y$ is performed beneath the center of the cavity at a depth $z = -20\,\mathrm{nm}\times(\lambda_0/737\,\mathrm{nm})$ which is scaled to the simulated resonance wavelength.
The target depth of $-20$\,nm is chosen as it matches the average depth of the SiV centers in our sample.

The parameter search is implemented using the global optimization algorithm LIPO~\cite{malherbe2017lipo} interleaved with local, derivative-less optimization via a trust-region optimization algorithm.
Both of these steps are implemented by Python's \texttt{dlib} package~\cite{king2009dlib}.
At a high level, the LIPO global search utilizes sparse samples of the parameter space to construct an upper bound on the objective function $\eta$ from which potential global maxima may be determined.
The LIPO algorithm quickly locates designs that are close to local maxima, however it is slow to converge on the precise local maxima.
Consequently, the local-trust region optimization is utilized to quickly estimate local maxima without requiring evaluation of the objective function gradient.

While this search algorithm greatly reduces the number of objective function evaluations, each evaluation still requires a costly FDTD simulation of the device design.
Thus, we utilize Flexcompute's Tidy3D FDTD simulation which enables a complete, moderate-resolution simulation of the device (and estimation of the objective function) within a few minutes.
During the search, simple bounds on the parameters are imposed to bias against designs that are impractical to fabricate.
Additionally, designs which do not have a resonance within 700--800\,nm are given an evaluation of 0.
We begin with the forward designs as the initial set of samples and run the search optimization for up to additional 500 evaluations.
Further evaluation is expected to yield only marginal improvements.

After completing the global search we rank order the simulations with respect to the objective function to identify the optimal designs.
Due to the local trust-region step, the best simulations are grouped into clusters centered on individual local maxima.
We identify the top 5 clusters and run the local trust-region optimization again on each cluster to estimate the true local maxima.
We pick the parameters corresponding to the largest local maxima as the device design.
Finally, we perform high-resolution FDTD simulation using both Tidy3D and Lumerical FDTD to verify the results.
The parameters are then scaled to shift the resonance to the target wavelength which establishes the final design.
The final design parameters and simulation results are tabulated in Table~\ref{tab:cavity_parameters}.

\begin{table}[h]
    \centering
    \begin{tabularx}{0.6\textwidth}{ X X X X X}
        \hline\hline \\
        \multicolumn{5}{l}{\textbf{Design parameters:}} \\
        \qquad$w$ & 402.8 nm & \qquad & $h$ & 182.8\,nm	\\
        \qquad$w_x$ & 66.3\,nm & \qquad & $w_y$ & 112.0\,nm	\\
        \qquad$a_{\mathrm{cav}}$ & 132.5\,nm & \qquad & $a_{\mathrm{mir}}$  & 140.1\,nm	\\
        \qquad$N_\mathrm{cav}$ & 12 & \qquad & $N_\mathrm{mir}$ & 25 \\
        \\
        \hline
        \\
        \multicolumn{5}{l}{\textbf{Simulated performance:}} \\
        \multicolumn{3}{l}{\qquad Intrinsic quality factor, $Q_i$} & \multicolumn{2}{l}{$1.75\times10^5$}\\
        \multicolumn{3}{l}{\qquad Normalized mode volume, $V$} & \multicolumn{2}{l}{$1.86\,(\lambda/n)^3$}\\
        \multicolumn{3}{l}{\qquad Field overlap, $\big|E_y(\vb x_a)/\max|\vb E|\big|^2$} & \multicolumn{2}{l}{$0.25$}\\
        \\
        \multicolumn{3}{l}{\qquad Objective function, $\eta$} & $5200$ & ($23500$)\\
        \multicolumn{3}{l}{\qquad Purcell enhancement, $F$} & 325 & ($1462$) \\
        \multicolumn{3}{l}{\qquad Cooperativity, $C$} & 23 & ($102$) \\
        \\
        \hline\hline
    \end{tabularx}
    \caption{
        \textbf{Optimized cavity design parameters.}
        The objective function, Purcell factor, and cooperativity are evaluated assuming an emitter placed at the center of the cavity, 20\,nm beneath the surface.
        The left value indicates the estimation assuming a fabrication limit $Q_\mathrm{fab}=5\times10^4$ while the right value in parethesis indicates the theoretical limit if $Q_\mathrm{fab}\to\infty$.
        Purcell enhancement is evaluated via $F = (3/4\pi^2) (Q/V) \big|E_y(\vb x_a)/\max|\vb E|\big|^2 \cos(35^\circ) = (3/4\pi^2) \eta \cos(35^\circ)$, where the cosine term originates from the $\langle111\rangle$ orientation of the SiV.
        Cooperativity is estimated via $C = \eta_\mathrm{QE}\eta_\mathrm{DWF} F$ where $\eta_\mathrm{QE} \approx 0.1$ and $\eta_\mathrm{DWF} = 0.7$ is the SiV quantum efficiency and Debye-Waller factor respectively.
    }
    \label{tab:cavity_parameters}
\end{table}

\subsubsection{Coupling optimization}
Finally we optimize the coupling of the cavity mode to the waveguide busses.
For the mirror-coupled geometry, this simply amounts to sweeping the number of mirror holes $N_\mathrm{mir}$ on either end.
For the side-coupled geometry we keep $N_\mathrm{mir}=25$ and instead sweep the size and position of an external coupling waveguide.
To optimize the coupling waveguide we first sweep the width $w_\mathrm{wg}$ of the adjacent GaP waveguide of identical height to the cavity $h$ while keeping a constant coupling distance $d_c$.
Using Lumerical's variational FDTD (varFDTD), we calculate the transmission while monitoring the resonance to determine the waveguide width which minimizes the loaded $Q$ factor.
Since varFDTD overestimates the intrinsic $Q$ factor, this width corresponds to the maximum waveguide-cavity coupling rate.
Physically, this corresponds to when the waveguide modes at the target frequency are phase-matched to the cavity field (which has wavevector $k_x \approx \pi/a_\mathrm{cav}$).
The coupling rate is then controlled by modulating the coupling distance $d_c$.

The two-port or ``add-drop'' geometry of either design does not support a zero in the transmission spectrum (i.e. critical coupling).
Thus, it is instead optimal to have as large of a coupling rate as possible to minimize the effect of the intrinsic losses.
However, increasingly loading the cavity in this manner will diminish the the total $Q$ factor and, in turn, the achievable cooperativity.
Moreover, fabrication losses will diminish the total $Q$ factor further, resulting in an even higher coupling rate being required.
To balance these effects and ensure visibility of the resonances, we simulate these designs for wide ranges of $d_c$ and $N_\mathrm{mir}$ to sweep from the far over-coupled to the far under-coupled regimes.
In practice this results in mirror numbers $N_\mathrm{mir} = 4$--$13$ and coupling distances $d_c = 50$--$140$\,nm.

\clearpage
\subsection{Device fabrication}
\begin{figure}[h]
    \centering
    \includegraphics{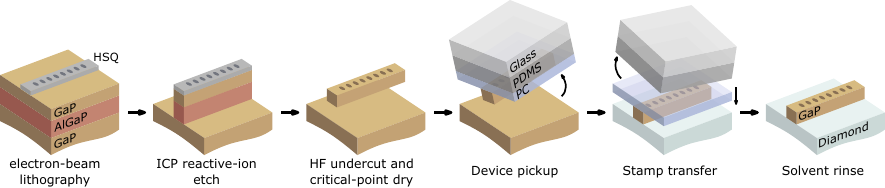}
    \caption{\textbf{Fabrication process.}
        An illustration of the fabrication process steps.
    }
    \label{fig:si:fabrication_process}
\end{figure}

Key steps in the fabricaiton process are outlined in Fig.~\ref{fig:si:fabrication_process}.
Additional details follow.

\subsubsection{Diamond sample preparation}
The diamond sample used in this study is the same sample used in our prior work Ref.~\cite{chakravarthi2023stamp}.
The sample is an electronic-grade, chemical-vapor-deposition diamond (Element Six) with N~$<$~5\,ppb, B~$<$~1\,ppb, and a surface orientation along the $(100)$ plane.
The edges of the sample are cut along the $\langle110\rangle$ axes of the diamond crystal.
The diamond was implanted with silicon ions accelerated to 12\,keV at an angle of 7$^\circ$ from the surface normal and a dose of $5\times10^{11}$\,ions/cm$^{2}$.
After implantation the sample was vacuum annealed at 1200\,$^\circ$C for 2 hours to form SiV centers and heal the implantation damage via vacancy diffusion.
Assuming a 1\% formation efficiency, the final SiV density is approximately $50$\,SiV/\textmu m$^2$.
SRIM Monte Carlo simulation of the implantation conditions indicates that the SiV are formed at depths with a roughly Gaussian distribution of mean $20$\,nm and standard deviation $6.5$\,nm (main text Fig.~1).
Previously stamped devices on the diamond were removed via mechanical abrasion and sonication in acetone, followed by rinses in isopropyl alcohol (IPA) and deionized (DI) water.

Prior to the later stamping step the diamond is again physically scrubbed and then sonicated in acetone prior to IPA and DI water rinses.
The sample is then dried with pressurized nitrogen gas.
It is then chemically cleaned in a boiling mixture of equal parts sulfuric, nitric, and perchloric acid (``tri-acid clean'') which removes any invisible organic contaminants.

\subsubsection{GaP sample preparation}
Approximately 200-nm-thick GaP membranes are grown on a 200-nm $\mathrm{Al_{0.8}Ga_{0.2}P}$ sacrificial layer on a quarter of a GaP(001) substrate using gas-source molecular beam epitaxy (GSMBE).
The sample was cleaved into $1$-cm wide chips for separate processing.
A given chip was cleaned with a series of solvent baths in acetone, IPA, and DI water (hereafter referred to as a ``standard solvent rinse'') then dried with pressurized nitrogen and a hot-plate dehydration bake at $150$\,$^\circ$C for 5 minutes.
Afterwards, an approximately $5$-nm layer of SiO$_2$ was deposited using a low-temperature (125\,$^\circ$C) PECVD process (SPTS) to assist in electron-beam resist adhesion.
The chips were then further cleaved into approximately $5$\,mm by $10$\,mm pieces which are used to write devices.

\subsubsection{Pattern preparation}
We pattern the devices using electron-beam lithography (EBL).
An individual pattern of devices consists of a $10\times4$ array of individual nanobeam devices.
The left two columns are identical and consist of side-coupled devices with coupling distances $d_c$ increasing in steps of $10$\,nm from $50$\,nm on the top row to $140$\,nm in the bottom row.
Likewise, the right two columns are identical and consist of mirror-coupled devices with increasing numbers of mirror holes $N_\mathrm{cav}$ increasing in steps of $1$ from $4$ to $13$.
Consequently going down the rows corresponds to a decreasing coupling rate and higher $Q$ factors, but lower contrast of the resonance.
The coupling waveguides of each device feed into/out of elliptical free-space grating couplers which are designed to minimize the Fabry-Perot oscillations in the waveguide while also yielding high vertical transmission (simulated vertical transmission of $15$\% over $700$--$800$\,nm, Supplemental 2B).
The cavity structure, waveguide, and gratings are embedded in a frame which is connected at 4 pinch-points to a larger rectangular cutout support frame.
The support frame holds the devices during the later undercut step and the sample breaks at the pinch points during the stamp transfer step.

Stochastic variations of the EBL process between fabrication runs results in some uncertainty in the final device dimensions.
To ensure that some devices are close to the intended design geometry and yield resonances within the target wavelength range we fabricate the device grid pattern in a $5\times7$ array where the nominal EBL dosage and hole-size is varied along the rows and columns respectively.
This yields shifts in the average resonance wavelength on the order of $1$\,nm between adjacent patterns in the grid (variations of the resonances within a pattern is itself within a few nanometers).
Finally, we fabricate three of these $5\times7$ grids where the design parameters $(w,w_x,w_y,a_\mathrm{cav},a_\mathrm{mir})$ are scaled to $100$\%, $101.5$\%, and $103$\% of the optimized design values to account for larger systematic variations.
This yields shifts in the average resonance wavelength on the order of $10$\,nm.

\subsubsection{Electron-beam lithography}
Individual GaP-on-AlGaP chips are sonicated for 5 minutes in acetone, followed by a standard solvent rinse, pressurized nitrogen drying and a 10 minute hotplate dehydration bake at 150\,$^\circ$C.
After cooling to room temperature the chip is spun with hydrogen silsesquioxane (HSQ) resist in the form of 6\% H-SiQ at $6000$\,rpm for $45$\,s which should yield a resist thickness of $100$--$150$\,nm.
The chip then soft baked at $80$\,$^\circ$C for $4$\,min.
Finally, an approximately 50-nm thick layer of DisCharge H20 x2 (DisChem) anti-charging agent is spun at $3000$\,rpm for $60$\,s to assist in the prevention of charging during the EBL write.

EBL is performed in a JEOL JBX-6300FS using the high-resolution (100\,keV) mode and
a current of 1\,nA.
This step patterns the devices and a surrounding support frame into the HSQ.
After the EBL write is completed, the sample is rinsed under running DI water for 30\,s to remove the DisCharge layer, after which it is dried with pressurized nitrogen.
The resist is then developed by a $4$-min soak in 25\% tetramethylammonium hydroxide (TMAH) and subsequent rinse in running DI water for 30\,s.
The sample is transported in DI water to a solvent bench where it can be rinsed in flowing IPA for 30\,s to remove any remaining HSQ residue.
The sample is then rinsed in DI water before drying with pressurized nitrogen.

\subsubsection{Device plasma etch}
The device pattern is transferred from the HSQ to the GaP layer via an inductively coupled plasma reactive-ion etch (ICP RIE) process.
The process is performed in an Oxford Instruments PlasmaLab 100 (ICP-180) and uses a 6:1:1 mixture of Ar:Cl$_2$:N$_2$ gasses with a chamber pressure of 3\,mTorr.
The etch is performed in three cycles of 30-s etch steps with 30-s rest steps to avoid heating.
Over 400\,nm of material is etched in bulk indicating an etch rate of about 4.5\,nm/s. 
This over etches the GaP and AlGaP layers which is useful in the later undercut steps.

\subsubsection{HF undercut and critical-point drying}
The chip is loaded into a Teflon sample holder and submerged in a 1.5\% hydroflouric acid (HF) wet etch for 130\,s.
The HF selectively targets the AlGaP layer yielding an approximately 2-3\,\textmu m undercut which completely releases the devices while maintaining AlGaP under the support frame, suspending the devices.
The chip is then transferred to a DI water bath without allowing it to dry (which would collapse the devices) and the water is drained and refilled 5 times while keeping the chip submerged to wash off any remaining HF.

A small amount of IPA is added to the DI water and submerged chip which reduces the surface tension and allows the devices to remain coated in the DI-IPA mixture while it is transferred to an IPA bath.
The sample is soaked in the IPA bath for 5\,min before being transferred to a bath of ultra-pure IPA for a 10-min soak.
Two more ultra-pure IPA bath soaks are performed to completely remove any remaining water.
The device is then transferred to a critical-point dryer (Tousimis 931) in which liquid CO$_2$ purges the IPA before being brought to the critical point where it is boiled off without damaging the devices.
The devices are now freely standing and can be measured.

\subsubsection{Stamp transfer}
The suspended devices are characterized using optical transmission spectroscopy to locate the resonance frequencies of the fundamental mode and determine the $Q$ factor.
The resonance frequencies are expected to redshift by approximately $10$\,nm when the device is transferred to the diamond substrate.
When the devices are cooled to cryogenic temperatures the resonances are expected to blueshift by approximately $5$\,nm.
The devices can be redshfited \textit{in situ} by several nanometers via gas deposition tuning so we want device resonances to be in the range of $734$--$737$\,nm.
Altogether, this indicates that devices with resonances at $729$--$732$\,nm should be considered for stamp transfer.
Fortunately our large array of devices support numerous sets of devices with resonances in this wavelength range.
The corresponding patterns (in the $3\times5\times7$ grid) are then flagged for stamp transfers.

\begin{figure}
    \centering
    \includegraphics{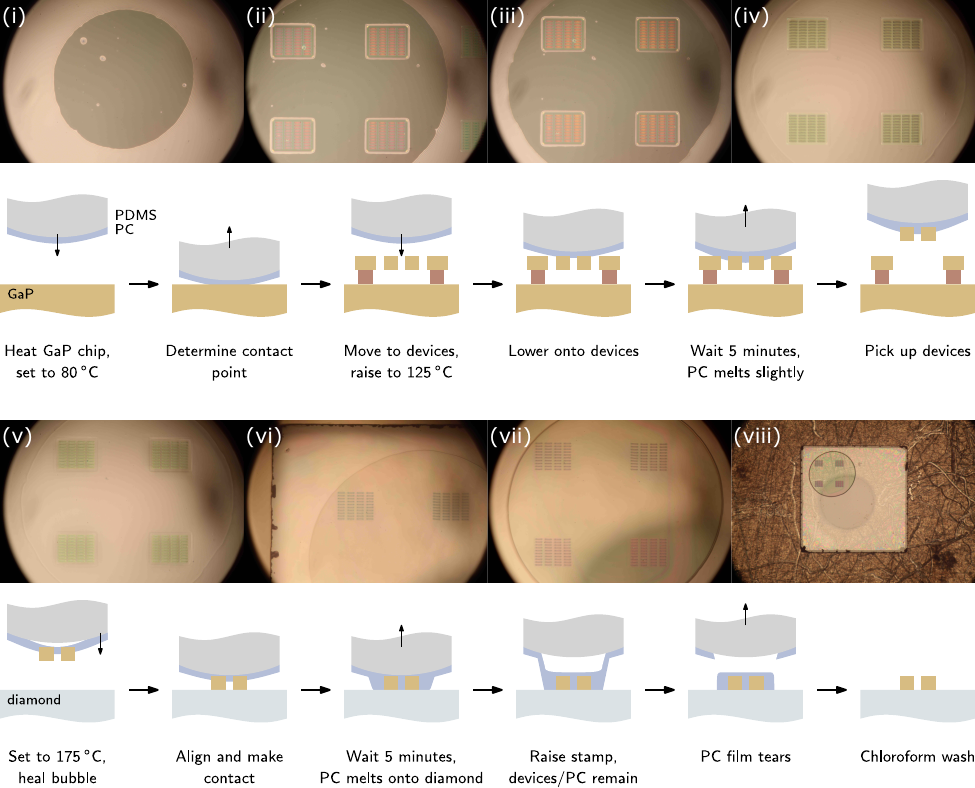}
    \caption{\textbf{Device stamping.}
        \textbf{Stamp transfer process.}
        (i) Locating the contact point of the stamp.
        (ii) Initial contact with the devices.
        (iii) After 5 minutes the color changes slightly to a more gold color.
        (iv) Empty device frames after picking up the devices.
        (v) Devices on the stamp above the diamond after healing any bubbles.
        (vi) Initial contact of the stamp with the diamond.
        (vii) PC film and transferred devices after releasing from the stamp.
        (viii) Wider view of the diamond after stamping (before later stamps and chloroform wash).
    }
    \label{fig:si:stamping}
\end{figure}

An illustration of the stamp transfer process and representative microscope images are shown in Fig.~\ref{fig:si:stamping}.
Stamp transfers are performed in a custom-built stamp transfer microscope with temperature and pressure control utilized for a variety of 2-d materials processes.
Stamps are formed from a glass substrate with a domed PDMS puck and thin layer of polycarbonate (PC) film stretched over top.
The dome curvature of the PDMS puck enables fine control over the contact area of the stamp such that single patterns of the $3\times5\times7$ grid can be stamped individually.

The stamp is then lowered onto one or more patterns on the chip, which is heated to 125\,$^\circ$C.
At this temperature the PC film is effectively an extremely viscous fluid and can flow over the pattern to improve the adhesion to the individual devices.
Over the course of a few minutes a visible change in the color of the devices is observed due to thin-film interference as the PC flows over the devices.
The stamp is then slowly raised, taking care to avoid separating the PC from the PDMS puck.
As the stamp raises, the devices are broken free from the support frame at the pinch points and remain stuck to the PC.

The stamp can then be moved over the diamond (which has been cleaned with solvent and the tri-acid clean) to begin the stamp transfer process.
The devices are aligned to the $\langle110\rangle$ edges of the diamond.
The position of the devices on the diamond surface can be placed with micron-scale precision.
The diamond is heated to 175\,$^\circ$C and the stamp with the devices is lowered onto the desired location.
At this temperature the PC melts completely, bonding to the diamond surface.
As the stamp is raised, the PC in contact with the diamond remains stuck to the surface forming a bubble in the PC film.
The stamp can then be raise further causing the film to tear, releasing the PC film and devices on the diamond.
This process can be repeated many times to transfer all devices of interest onto a single chip.

After all devices of interest are transferred, the diamond is allowed to rest for 24\,hours before soaking for 6\,hours in chloroform at room temperature to dissolve the remaining PC film.
Finally, the devices are rinsed with acetone, IPA, and DI water before drying with pressurized nitrogen.
This completes the fabrication process.
We successfully transfer $637$ of $640$ devices onto the diamond corresponding to a $99.5$\% yield.

\clearpage
\section{Theoretical modeling}
\subsection{Cavity transmission} \label{sec:si:modeling:transmission}
The system at zero magnetic field can be modeled by a single two-level system spanned by $\{\ket{g},\ket{e}\}$ coupled to a single cavity mode $\hat a$.
This system is described by the Jaynes-Cummings Hamiltonian
\begin{equation}
    \hat H = \hbar\omega_0 \hat a^\dagger \hat a
    + \hbar\omega_1 \hat \sigma_{ee}
    - \hbar g \Big( \hat a\hat\sigma_{ge}^\dagger + \hat a^\dagger\hat\sigma_{ge} \Big)
\end{equation}
where $\sigma_{ij} = \dyad{i}{j}$ for $i,j\in\{g,e\}$, $\omega_0$ and $\omega_1$ are the cavity and atomic transition frequencies, and $g$ is the single-photon Rabi frequency.
The system is subject to input and damping via a number of bosonic reservoirs with associated rates and jump operators as:
\begin{itemize}
    \item Intrinsic cavity loss: $(\kappa_i,\hat a)$,
    \item Coupling cavity loss (two identical channels for each set of waveguide modes): $(\kappa_c,\hat a)$,
    \item SiV excited state decay: $(\gamma,\hat\sigma_{ge})$,
    \item SiV excited state pure dephasing: $(\gamma_d,\hat\sigma_{ee})$.
\end{itemize}
This is incorporated into the model via the input-output formalism~\cite{gardiner1985heisenberglangevin} which yields the Heisenberg-Langevin (H-L) equations
\begin{subequations}
    \begin{align}
        \dv{a}{t} &= - \Big(i\omega_0 + \frac{\kappa_i + 2\kappa_c}{2}\Big) a 
        +ig\sigma_{ge}
        - \sqrt{\kappa_c} s_+, \\
        \label{sec2:zero-field-hl-eqns-2}
        \dv{\sigma_{ge}}{t} &= - \Big(i\omega_1 + \frac{\gamma + \gamma_d}{2}\Big)\sigma_{ge} 
        +ig a \big(\sigma_{gg} - \sigma_{ee}\big), \\
        s_{j-} &=  s_{j+} + \sqrt{\kappa_c}  a, \\
        f_- & = \sqrt{\gamma} \sigma_{ge},
    \end{align}
\end{subequations}
where the dynamical variables $a(t) = \ev{\hat a}$ and so on are the expectation values of the system operators (noise terms have been dropped accordingly).
These equations are truncated at linear operator expectation values via a cumulant expansion~\cite{kubo1962cumulant,sanchez2020cumulant} such that the correlations are ignored (i.e.~$\ev{\hat a \hat \sigma_{gg}} \approx a \sigma_{gg})$.
The fields $s_{j\pm}$ for $j\in\{1,2\}$ are the ensemble-averaged input/output fields of the two waveguide baths and $f_-$ corresponds to the average field of the free-space modes through which the SiV can decay and is directly proportional to the SiV florescence.

We assume that the driving $s_{j+}(t)$ is sufficiently weak such that it does not on-average modify the SiV populations, i.e.~$\sigma_{ee} = 0$ and $\sigma_{gg} \approx 1$.
The equations are then linear and can be solved via a Fourier transform $a(t)\to A(\omega)$ and $\dot a(t) \to -i\omega A(\omega)$.
We determine the scattering matrix elements $S_{jk}(\omega) = S_{j-}(\omega)/S_{k+}(\omega)$ as
\begin{subequations}
    \begin{align}
        \mathrm{thru:}\qquad
        S_{11}(\omega) = S_{22}(\omega) &= 
            \frac{
                \Big(i\delta - \kappa_i/2\Big)
                \Big(i(\delta - \Delta) - (\gamma+\gamma_d)/2\Big)
                +
                g^2
            }{
                \Big(i\delta - (\kappa_i + 2\kappa_c)/2\Big)
                \Big(i(\delta - \Delta) - (\gamma+\gamma_d)/2\Big)
                +
                g^2
            },  \\
        \mathrm{drop:}\qquad
        S_{21}(\omega) = S_{12}(\omega) &= 
            \frac{
                \kappa_c
                \Big(i(\delta - \Delta) - (\gamma+\gamma_d)/2\Big)
            }{
                \Big(i\delta - (\kappa_i + 2\kappa_c)/2\Big)
                \Big(i(\delta - \Delta) - (\gamma+\gamma_d)/2\Big)
                +
                g^2
            },
    \end{align}
\end{subequations}
where $\delta = \omega - \omega_0$ is the drive-cavity detuning and $\Delta = \omega_1 - \omega_0$ is the SiV-cavity detuning.
For side-coupled devices in the lossy-cavity regime $\kappa\gg g,\gamma$, the transmission $T_\mathrm{thru}(\omega) = |S_{11}|^2$ takes the form of a Lorentzian dip at the cavity resonance with approximate width $\kappa$ and narrow transparency window at the SiV resonance frequency.
This corresponds to the well-known dipole-induced transparency (DIT) effect~\cite{waks2006dit}.
Alternatively for mirror-coupled devices, the transmission is instead $T_\mathrm{drop}(\omega) = |S_{21}(\omega)|^2$ which corresponds to a Lorentzian peak and narrow scattering window at the SiV.
For simplicity, this latter effect is also referred to as DIT.

The fluorescence of the SiV center can be modeled by the scattering of light into the free space modes given by
\begin{equation}
    F_-(\omega) = 
    \frac{
        -ig\sqrt{\kappa_c\gamma} S_{1+}(\omega)
    }{
        \Big(i\delta - (\kappa_i + 2\kappa_c)/2\Big)
        \Big(i(\delta - \Delta) - (\gamma+\gamma_d)/2\Big)
        +
        g^2
    }.
\end{equation}
In the lossy-cavity regime ($\kappa\gg g,\gamma$), the fluorescence $F_-(\omega)$ is sharply peaked at the SiV transition frequency $\delta = \Delta$ (the cavity and SiV are only weakly hybridized).
Consequently, we may approximate $(i\delta - \kappa/2)\approx(i\Delta-\kappa/2)$.
Factoring this term out we can write the fluorescence as
\begin{equation}
    F_-(\omega) = \frac{-ig\sqrt{\kappa_c\gamma} S_{1+}(\omega)}{\Big(i\Delta - (\kappa_i + 2\kappa_c)/2\Big)}
    \frac{1}{\Big(i(\delta - \Delta_\mathrm{eff}) - \gamma_\mathrm{eff}/2\Big)}
\end{equation}
where the first factor describes the cavity excitation efficiency and
\begin{subequations}
    \begin{align}
        \Delta_\mathrm{eff} &= \Delta \bigg( 1 + \frac{g^2}{\Delta^2 + \kappa^2/4} \bigg), \\
        \gamma_\mathrm{eff} &= \gamma + \gamma_d + \frac{4g^2}{\kappa}\frac{\kappa^2/4}{\Delta^2 + \kappa^2/4}
    \end{align}
\end{subequations}
are the effective detuning and linewidth respectively.
The modification to the detuning corresponds to the cavity Lamb shift and is generally negligible in the parameter regime of interest ($\kappa\gg g > \gamma$).
In contrast, the modification to the linewidth corresponds to Purcell enhancement~\cite{chakravarthi2023stamp} and can result in broadening of the linewidth on the scale of $\gamma$.
Thus, the SiV fluorescence intensity $I(\omega) = |F_-(\omega)|^2$ takes the form of a Lorentzian peak at $\delta = \Delta_\mathrm{eff}$ and with linewidth $\gamma_\mathrm{eff}$.

This analysis presumes that the SiV can be treated as a true two-level system.
However, the SiV possesses an additional orbital level in both the ground and excited state manifolds which can trap population.
When the SiV is trapped in such a state, the coupling term in Eq.~\eqref{sec2:zero-field-hl-eqns-2} vanishes (since $\sigma_{gg} = \sigma_{ee} = 0$) resulting in scattering amplitudes matching the bare add-drop cavity
\begin{subequations}
    \begin{align}
        \mathrm{thru:}&&
        S_{11}^\mathrm{dark}(\omega) = S_{22}^\mathrm{dark}(\omega) =& 
            \frac{
                \big(i\delta - \kappa_i/2\big)
            }{
                \big(i\delta - (\kappa_i + 2\kappa_c)/2\big)
            },  &\\
        \mathrm{drop:}&&
        S_{21}^\mathrm{dark}(\omega) = S_{12}^\mathrm{dark}(\omega) =& 
            \frac{
                \kappa_c
            }{
                \big(i\delta - (\kappa_i + 2\kappa_c)/2\big)
            }, &\\
        \mathrm{fluorescence:}&& F_-^\mathrm{dark}(\omega) =& 0.&
    \end{align}
\end{subequations}
Nevertheless these additional levels can be neglected insofar as (1) any optical transitions involving the additional levels are far detuned from the transition of interest, and (2) the thermalization of the ground states is sufficiently fast compared to the driving.
The SiV satisfies both of these conditions as the ground state orbital splittings are large ($\Delta E/h \approx 50$\,GHz) and thermalize quickly with lifetime $T_1^\mathrm{orbit} \approx 38$\,ns at $4$\,K~\cite{jahnke2015sivelectronphonon} (which is much faster than typical excitation fluences of $10^6$\,photons/s).
Assuming such thermalization occurs before each scattering event, the scattered fields are then a classical ensemble average with
\begin{subequations}
    \begin{align}
        T(\omega) &= p_g|S_{ij}(\omega)|^2 + (1-p_g) |S_{ij}^\mathrm{dark}(\omega)|^2, \\
        I(\omega) &= p_g|F_-(\omega)|^2 + (1-p_g) |F_-^\mathrm{dark}(\omega)|^2,
    \end{align}
\end{subequations}
where $p_g$ is the thermal-average population in the bright state $\ket{g}$ which obeys Boltzmann statistics with $p_g = (1+\exp(-\Delta E/k_B T))^{-1}$.
For an average SiV with ground-state splitting $\Delta E/h = 50$\,GHz, one has $p_g \approx 0.65$ at $T=4$\,K.
With respect to the SiV fluorescence, this simply results in a diminished intensity.
The transmission spectrum is modified such that the DIT peak contrast is diminished proportionally giving the appearance of a reduced cooperativity.
However, in the lossy-cavity regime ($\kappa\gg g,\gamma$) the dispersion of the background cavity transmission is largely negligible over the DIT signal bandwidth and so is indistinguishable from the background signal.
As such, the transmission can be directly fit to the idealized two-level model $T_\mathrm{thru}(\omega)$ or $T_\mathrm{drop}(\omega)$, however such fits systematically underestimate the coupling rate $g$ and thus yield lower bounds on the expected cooperativity.

\clearpage
\subsection{Free-space grating couplers}
We design elliptical free-space grating couplers to couple light into and out of the devices.
The grating design consists of elliptical slots with quasi-periodic spacing, positioned such that one focus of the ellipse is centered on the input waveguide and the other focus is located on a line angled at an angle $\phi$ from the waveguide axis.
This causes any light incident from the waveguide that is not scattered upward to be directed towards the off-axis focus and scattered at an oblique angle out of the GaP layer, rather than being reflected back into the waveguide.
This is necessary to eliminate the low-finesse Fabry-Perot oscillations between the input/output grating couplers which diminishes the signal quality.

The arcs of the grating slots are defined by the polar equation
\begin{equation*}
    r(\theta) = \frac{a(1-e^2)}{1-e\cos(\theta-\phi)},
\end{equation*}
where $a$ defines the major axis length and $e$ defines the ellipse eccentricity.
The period $A$ and duty cycle $d$ of the grating are defined with respect to $a$ such that the start of the $(n+1)$-th grating period has corresponding major axis length $a_{n+1} = a_{n} + A$ and the slot begins at $a_{n} + d\cdot A$.
Our gratings have parameters as defined in Table~\ref{tab:si:grating_parameters}.

\begin{table}[h]
    \centering
    \begin{tabularx}{0.25\textwidth}{ Xl }
        \hline\hline \\
        Grating angle           & $45^\circ$ \\
        Periods                 & 4 \\
        Starting arc $a_0$      & 3\,\textmu m \\
        Periodicity $A$         & 0.5\,\textmu m\\
        Duty cycle $d$          & 40\% \\
        Eccentricity $e$        & 0.20 \\
        Axis angle $\phi$       & $-30^\circ$\\
        \\
        \hline\hline
    \end{tabularx}
    \caption{\textbf{Elliptical grating coupler parameters.}}
    \label{tab:si:grating_parameters}
\end{table}

\begin{figure}[h]
    \centering
    \includegraphics{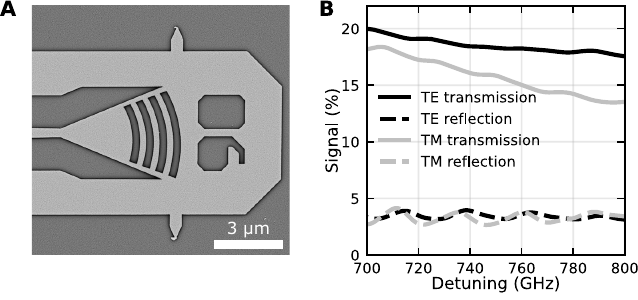}
    \caption{
        \textbf{Elliptical free-space grating coupler.}
        (\textbf{A}) SEM image of a grating coupler after transfer to diamond.
        The texture is due to a gold/palladium coating for imaging.
        (\textbf{B}) Transmission and reflection for TE/TM light injected from the waveguide (left side in (\textbf{A})).
        Input coupling is likely to be significantly reduced due to the small NA of the grating.
    }
    \label{fig:si:grating}
\end{figure}

The grating parameters are chosen to achieve a balance between low reflectivity back into the waveguide with high and broadband transmission upward into a relatively tight angular distribution.
The reflection and transmission are shown in Fig.~\ref{fig:si:grating}B as simulated in Lumerical FDTD.
We expect about 20\% upward transmission over the entire range of interest.
Note that the input coupling is likely much lower due to the relatively small numerical aperture (NA) of the grating itself.
High eccentricity causes the upward light to be scattered at large angles from the surface normal which may diminish collection efficiency due to the finite NA of the collection objective.
Finally, we note that collection via near-field fiber coupling may yield significant improvements in input/output coupling for long-term scalability.

\clearpage
\section{Data collection and analysis}
\subsection{Experimental setup}
\begin{figure}[h]
    \centering
    \includegraphics{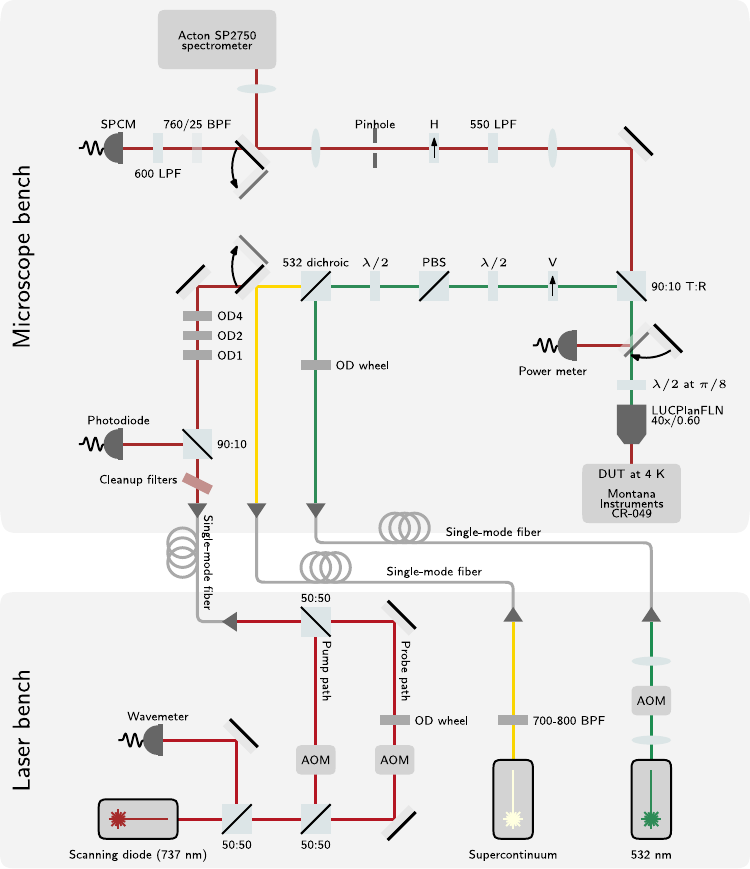}
    \caption{\textbf{Schematic of the microscope setup.}
        Resonant excitation is performed with the scanning diode laser (Sacher Lasertechnik Lion).
        Broadband transmission measurements are performed using the supercontinuum laser (Fianium Whitelase).
        Off-resonant excitation is performed with a 532-nm diode laser.
        All electrical I/O is performed by a National Instruments PCIe-6323 DAQ board. 
        Fast timing resolution is obtained by a Swabian Instruments Time Tagger Ultra. 
    }
    \label{fig:si:microscope}
\end{figure}

\clearpage
\subsection{Gas tuning}
\begin{figure}[h]
    \centering
    \includegraphics{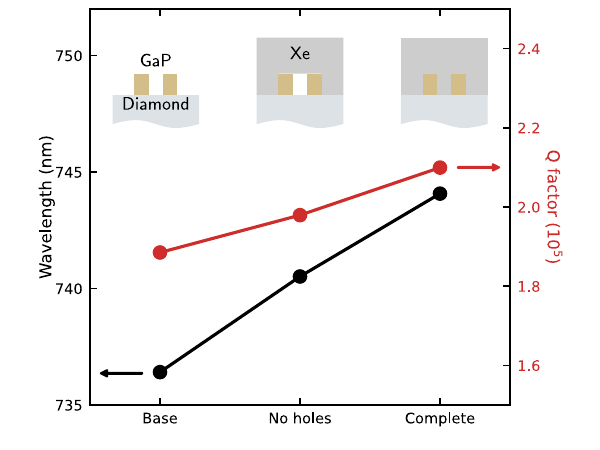}
    \caption{\textbf{Xenon gas tuning.}
        Resonance wavelength and quality factor for three different xenon gas amounts.
        ``Base'' corresponds to the bare system without any xenon.
        ``No holes'' describes the case where xenon covers everything but does not fill the holes.
        ``Complete'' describes the case where the holes are filled.
    }
    \label{fig:si:gas_tuning}
\end{figure}

Small amounts of xenon gas are injected into the cryostat chamber to controllably redshift the resonances as it condenses onto the sample.
We continuously monitor the resonance as the tuning occurs.
The xenon gas, when frozen, is modeled as having a refractive index of $1.49$.
The structure is simulated with various amounts of xenon in Lumerical FDTD, with the resonance position and quality factor plotted in Fig.~\ref{fig:si:gas_tuning}.
Our limited gas tuning range of about 2\,nm is attributed to the xenon gas being unable to make it into the holes of the photonic crystal as a result of our cryostat and gas line geometry.

\clearpage
\subsection{Cavity transmission spectroscopy}

\subsubsection{Broadband transmission spectroscopy}
Cavity transmission spectra are measured via a high-resolution grating spectrometer (Princeton Acton 2750).
Narrowband laser reflection measurements near 737\,nm are used to correlate the wavemeter frequency reading with the spectrometer wavelength.
Scanning laser measurements at zero field were performed using a Burleigh WA-1600 wavemeter, while the spin measurements used a HighFinesse WS7 wavemeter.

Individual transmission spectra are obtained by coupling a focused supercontinuum laser (Fianium Whitelase) into the input free-space grating coupler and sending the transmitted signal to the spectrometer.
Importantly, the input light aligned along $V$ (TE) is rotated to 45$^\circ$ by a half-wave-plate (HWP) above the objective such that it equally excites TE and TM modes.
Direct scatter that does not couple into the device is rotated back to $V$ and subsequently cross polarized with an output polarizer aligned along $H$.
Additional filtering of the direct scattering is achieved via spatially filtering through a pinhole.
Light that propagates through the device is projected into either TE or TM which are rotated to $\pm45^\circ$ after leaving the objective, which are both able to transmit through the output $H$ polarizer, albeit with an intensity reduced by $50\%$.

\begin{figure}[h]
    \centering
    \includegraphics{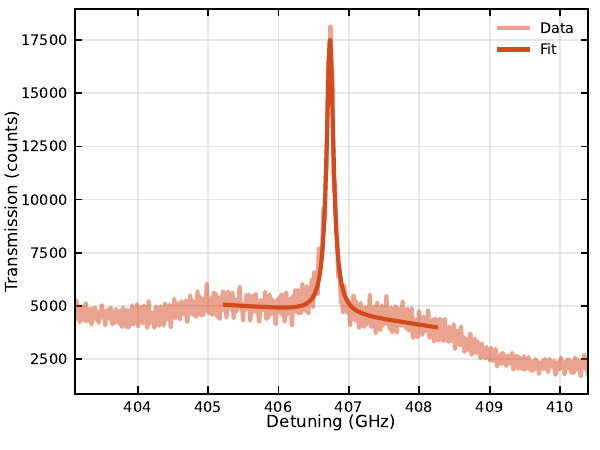}
    \caption{
        \textbf{Broadband cavity transmission.}
        Representative transmission with fit for the device presented in the main text.
    }
    \label{fig:si:broadband_tx}
\end{figure}

Consequently, the output signal contains three contributions: (1) the TE cavity transmission, (2) broadband TM device transmission, and (3) the heavily attenuated direct reflection (Fig.~\ref{fig:si:broadband_tx}).  
The baseline count rate of 590 counts (estimated from dark-count calibration measurements) is subtracted and the resulting spectrum is fit to the expected Lorentzian peak (TE contribution) with an arbitrary quadratic background (TM contribution), i.e.
\begin{equation} \label{eq:si:broadband_tx}
    y(\omega) = a \frac{(\kappa/2)^2}{(\omega-\omega_0)^2 + (\kappa/2)^2} + \Big(b_0 + b_1(\omega-\omega_0) + b_2(\omega-\omega_0)^2\Big),
\end{equation}
where $\kappa$ is the resonance linewidth, $\omega_0$ is the resonance frequency, and $b_j$ are coefficients for a Taylor expansion of the background in the vicinity of the resonance.
The fit is performed assuming shot-noise uncertainty at each of the data points.

Proper estimation of the relative intensity of the TE ($\sim a$) and TM ($\sim b_j$) contributions is important for determining the cooperativity, which is correlated to the transmission spectrum contrast at the SiV resonance.
Therefore, before each spectrum is taken, the input coupling and output collection are adjusted via the piezo stage to maximize the ratio of the TE/TM contributions without diminishing the overall count rate.
Estimation of the cooperativity assuming the maximal TE/TM ratio then represents a lower bound.

\subsubsection{DIT/PLE measurements}
At the start of each measurement session, the cavity resonance position is measured via a broadband transmission measurement described above.
The resonance position $\omega_0$ and linewidth $\kappa$ are determined from this measurement.
The relative contribution of the background and cavity transmission is also calibrated with this measurement.
The cavity and SiV centers are resonantly excited by a scanning diode laser (Sacher Lasertechnik Lion) which is focused through one of the input grating couplers.
Transmission measurements are performed by collecting the transmitted light using the same geometry as the broadband transmission measurements.
PLE measurements collect light out of the top of the cavity through a 760/25 band-pass filter and without the output polarizer and pinhole.
Typical excitation powers range between $0.5$--2\,nW before the objective which typically yields about a 100\,kcts/s at the SPCM (in transmission) when the cavity is driven near resonantly.
At large detunings, increased powers are used for PLE due to the reduced excitation efficiency.

\begin{figure}[h]
    \centering
    \includegraphics{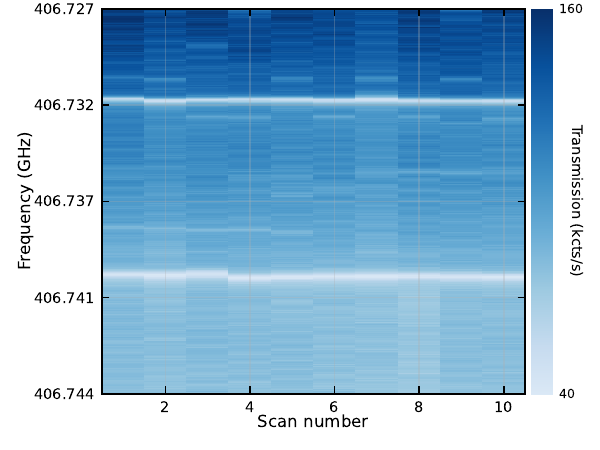}
    \caption{
        \textbf{DIT scans.}
        Representative set of narrowband transmission scans (this is the same dataset utilized to generate main text Fig.~3B).
        Each scan takes 40\,s, with a 3-s scan in the opposite direction to reset the frequency and 10-s repump step.
        The two DIT dips shown correspond to SiV 1 and 2 (top and bottom respectively).
    }
    \label{fig:si:dit_scans}
\end{figure}

Most measurements for fitting are performed using comparatively narrow-bandwidth scans covering only about 8--20\,GHz (Fig.~\ref{fig:si:dit_scans}).
The laser intensity dispersion in this regime is relatively linear and we utilize a fast photodiode to track the input power.
We collect the spectrum by repeatedly scanning the laser over the fixed bandwidth.
Typical scan durations last from 10--40\,s in one direction with a faster (typically 1--5\,s) backward scan to the starting position.
A few-second 532-nm laser repump pulse is interleaved after each forward/backward sweep to initialize the SiV into the negatively charge state and to allow for the laser to stabilize.

The unit-normalized cavity transmission is given by (Sec.~\ref{sec:si:modeling:transmission})
\begin{equation*}
    T(\delta) = \abs{
        \frac{
            (\kappa/2)(i(\delta-\Delta) - \gamma/2)
        }{
            (i\delta - \kappa/2)(i(\delta-\Delta) - \gamma/2) + g^2
        }
    }^2,
\end{equation*}
where $\delta = \omega - \omega_0$ is the drive detuning and $\Delta = \omega_1 - \omega_0$ is the SiV-cavity detuning.
Note that the cavity position $\omega_0$ and linewidth $\kappa$ are determined by the broadband spectroscopy measurements.
The actual measured signal contains contributions from the TM transmission and scattered background.
Additionally, there is a weak Fabry-Perot oscillation which we include as a multiplicative second-order polynomial.
The fit model for the DIT signal is then
\begin{equation}
    y(\delta) = \bigg[
        T(\delta) 
        + 
        \frac{1}{a}\big( b_0 + b_1 \delta + b_2 \delta^2 \big)
    \bigg]
    \cdot
    \big( f_0 + f_1 (\delta - \Delta) + f_2 (\delta - \Delta)^2 \big),
\end{equation}
where $a$, $b_j$ are determined from the earlier fit Eq.~\ref{eq:si:broadband_tx} and $f_j$ are constants for the Fabry-Perot Taylor expansion.
We assume shot-noise at each point in the spectrum during the fit.
After fitting all of the scans in a given measurement sequence, we discard any scans with poor fits before estimating the parameters as the most-likely estimator.
For example, given $N$ samples of a parameter $\mu_j$ with errors $\sigma_j$, the most-likely estimator is $\bar{\mu}\pm\bar{\sigma}$ where
\begin{align*}
    \bar{\mu} &= \frac{1}{\sum_j (1/\sigma_j)^2} \sum_j \frac{\mu_j}{\sigma_j^2}, \\  
    \bar{\sigma} &= \frac{1}{\sum_j (1/\sigma_j)^2} .
\end{align*}
These are the values reported in the main text.

\clearpage
\subsection{Estimation of loss and transmission in fabricated devices}
With respect to the device described in the main text, a simulation (Lumerical FDTD) of a device with an identical number of mirror and cavity holes ($N_{\mathrm{mir}}=6$, $N_{\mathrm{cav}}=12$) and the optimal design geometry yields a total $Q$ factor of $Q_{\mathrm{sim}} = 4250$.
The simulated $Q$ factor is assumed to be composed of design-limited intrinsic radiative losses $Q_{\mathrm{rad}}$ and extrinsic waveguide-coupling losses $Q_{\mathrm{c}}$ as
\begin{equation*}
    \frac{1}{Q_{\mathrm{sim}}} = \frac{1}{Q_{\mathrm{rad}}} + \frac{2}{Q_{\mathrm{c}}},
\end{equation*}
where the factor of $2$ indicates that $Q_\mathrm{c}$ corresponds to the coupling $Q$ factor for one of the symmetric waveguide ports.
From coupled-mode theory we expect the reflection (thru) and transmission (drop) signals on resonance to be given by
\begin{equation*}
    R_{\mathrm{sim}} = \frac{Q_{\mathrm{sim}}^2}{Q_{\mathrm{rad}}^2}, 
    \qquad\qquad
    T_{\mathrm{sim}} = \frac{4Q_{\mathrm{sim}}^2}{Q_{\mathrm{c}}^2}. 
\end{equation*}
The simulated transmission and reflection are given by $T_{\mathrm{sim}} = 0.90$ and $R_{\mathrm{sim}} = 0.01$ respectively.
The reflection amplitude is within the simulation noise floor and so we estimate the coupling and radiative $Q$ factors using the transmission value.
This yields an estimate of the radiative loss $Q_{\mathrm{rad}} = 83000$ and coupling loss $Q_{\mathrm{c}} = 8960$.

The measured device has a total $Q$ factor of approximately $Q_{\mathrm{exp}} = 3540$.
We assume that the simulated $Q_{\mathrm{rad}}$ is an upper bound and that $Q_{\mathrm{c}}$ is largely unaffected by imperfections in the realized device.
These assumptions are valid for small perturbations around the simulated optimal design.
Thus we attribute the reduction in $Q_{\mathrm{exp}}$ to an additional source of ``intrinsic'' fabrication losses $Q_\mathrm{fab}$ satisfying
\begin{equation*}
    \frac{1}{Q_{\mathrm{exp}}} = \frac{1}{Q_{\mathrm{sim}}} + \frac{1}{Q_{\mathrm{fab}}}.
\end{equation*}
From this we determine $Q_\mathrm{fab} = 21000$ which is roughly consistent with the highest measured $Q$ factors on the chip.
In principle, $Q_\mathrm{fab}$ represents all additional losses incurred by the imperfections in fabrication (roughness, material absorption, imperfect geometry).
The fabrication loss enters into the calculation as a reduced $Q_\mathrm{rad}$ and so we can write the estimated transmisison and reflection as
\begin{equation*}
    R_{\mathrm{exp}} = \frac{Q_{\mathrm{exp}}^2}{Q_{\mathrm{rad}}^2} \approx 0.002,  
    \qquad\qquad
    T_{\mathrm{exp}} = \frac{4Q_{\mathrm{exp}}^2}{Q_{\mathrm{c}}^2} \approx 0.62.
\end{equation*}
We can then extract the associated loss rates via $\kappa_j = \omega_0 / Q_j$ where $\omega_0/2\pi \approx 406.77$\,THz is the resonance frequency.
The results are summarized below in Table~\ref{tab:si:estimated_coupling}.

\begin{table}[h]
    \centering
    \begin{tabularx}{0.5\textwidth}{ X X X X X }
        \hline \hline 
        \\
        \quad$Q_\mathrm{exp}$ & 3540  & \qquad\qquad
            & $\kappa_\mathrm{exp}/2\pi$ & 115\,GHz \\
        \quad$Q_\mathrm{sim}$ & 4250  & \qquad\qquad
            & $\kappa_\mathrm{sim}/2\pi$ & 95.7\,GHz \\
        \\
        \hline
        \\
        \quad$Q_\mathrm{rad}$ & 83000  & \qquad\qquad
            & $\kappa_\mathrm{rad}/2\pi$ & 4.9\,GHz \\
        \quad$Q_\mathrm{c}$ & 8960  & \qquad\qquad
            & $\kappa_\mathrm{c}/2\pi$ & 45.4\,GHz \\
        \quad$Q_\mathrm{fab}$ & 21000  & \qquad\qquad
            & $\kappa_\mathrm{fab}/2\pi$ & 19.4\,GHz \\
        \\
        \hline\hline
    \end{tabularx}
    \caption{\textbf{Estimated $Q$ factors and loss rates for device in main text}}
    \label{tab:si:estimated_coupling}
\end{table}

\clearpage
\subsection{System efficiency}
We estimate the microscope efficiency by measuring the 737-nm laser power at various points in the system and correlating power measurements with photo-detection count rates at the SPCM (Excelitas SPCM-AQ4C) through calibrated neutral-density filters.
These values are tabulated in Tab.~\ref{tab:si:system_efficiency}.
The final item describing ``SPCM detection efficiency'' is directly measured.
We note that this is anomalously lower than the expected SPCM detection efficiency of roughly 60\% in this wavelength range.
The reduced efficiency, of roughly 35\%, is due in part to the fiber coupling loss but is largely of unknown origin.

Additionally, these calculations do not take into account the input/output coupling loss of the free-space grating couplers.
Prior simulations indicate an approximately 18\% output collection efficiency, resulting in only about 1\% of the photons transmitted through the cavity being collected.
These losses could be significantly reduced with fiber input/output coupling.

\begin{table}[h]
    \centering
    \begin{tabularx}{0.45\textwidth}{ Xl }
        \hline\hline \\
        Input (power meter to sample)           & $0.5$ \\
        Output optics                           & $0.725$ \\
        Output polarizer (for tx measurements)  & $0.5$ \\
        SPCM detection efficiency               & $0.207$ \\
        \\
        \hline
        \\
        Total collection efficiency (sample to SPCM)  & $0.075$ \\
        \\
        \hline\hline
    \end{tabularx}
    \caption{\textbf{Measured microscope efficiencies.}}
    \label{tab:si:system_efficiency}
\end{table}

\clearpage
\subsection{Additional device measurements}
\subsubsection{Post-stamping transmission measurements}
\begin{figure}[h]
    \centering
    \includegraphics{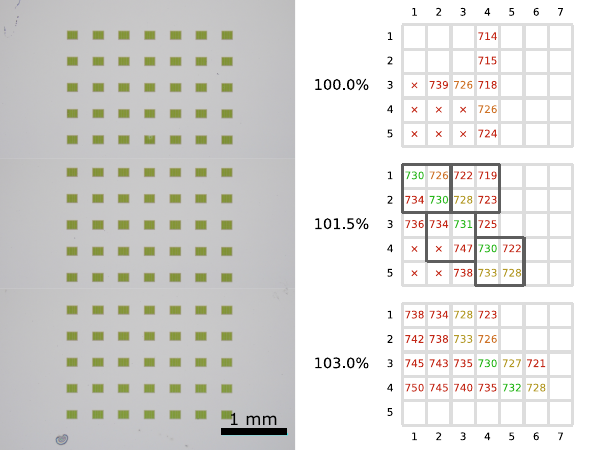}
    \caption{
        \textbf{Resonance wavelengths prior to stamping.}
        Composite optical image of the GaP chip after etching (left).
        Representative resonance wavelength for the devices in a given set prior to stamping.
        The colors qualitatively describe sets which have resonances near the target range (730--732\,nm).
        Empty cells were not measured.
        Cells marked with ``$\times$'' did not have clear resonances.
        The black squares correspond to the devices which were stamped.
    }
    \label{fig:si:prestamp_devices}
\end{figure}

We fabricate 4200 devices on a GaP-on-AlGaP chip using a standard EBL process.
The GaP chip, shown in Fig.~\ref{fig:si:prestamp_devices} after etching, has three $5\times 7$ arrays of devices indexed as $i \in \{1,2,3\}$.
Each of the three arrays, from top to bottom, corresponds to an overall scaling of the PhC parameters to $100.0$\%, $101.5$\%, and $103.0$\% of the optimal design values.
This is done to compensate for an observed blue shift in the zeroth order resonances in prior fabrication runs.
Stochastic variation of the EBL performance between different fabrication runs also produces noticeable effects on the final device performance.
To mitigate this effect, within each array, the base EBL dose and size of the PhC holes is modulated linearly along the rows $j \in \{1,...,5\}$ and columns $k \in \{1,...,7\}$ respectively.
A given device then is labeled by $(i,j,k)$.
The overall effect of these modulations is that the resonance redshifts as $i$, $j$ increases and $k$ decreases.

Only a small subset of the devices achieve resonances that can be tuned onto the SiV transition.
The frequency of the zeroth-order resonances of the suspended devices are shown in Fig.~\ref{fig:si:prestamp_devices}.
Not all sets were measured as their resonances could not be resolved, or they were expected to be shifted too far from the range of interest.
Resonances of the suspended devices are expected to redshift by about 10\,nm after stamping on to diamond, then blue shift by 5\,nm as the sample is cooled to 4\,K.
Consequently, the ideal resonance wavelengths are around 730--732\,nm.

\subsubsection{DIT spectra on other devices}
In the following, individual devices will be referred to as \texttt{set$(j,k)$dev$(r,c)$}, where $r$, $c$ indicates the row and column of the device within the transferred set $(j,k)$.
For example, the device presented in the main text corresponds to \texttt{set$(1,1)$dev$(3,3)$} (top left pattern, third row, third column).

After cooling to 4\,K, the transmission spectra of promising devices were measured to confirm that the resonance was within the tuning range of the SiV $C$ transition.
Of those that were, we selected a subset of devices whose resonances were well coupled enough to appear as high-contrast peaks/dips in the broadband transmission spectra.
The initial resonance position (immediately after the cooldown), and approximate quality factor for these devices is tabulated below in Tab.~\ref{tab:si:dit_devices}.
These devices were examined for DIT signal as they were tuned to the SiV $C$ line.
We measure scanning transmission spectra on 9 of the 12 devices of interest.
Of the 9 measured devices, 2 did not appear to have any clear DIT signal, 3 had moderate DIT signal and were not pursued further, and 4 had strong DIT signal.
Due to the considerable time investment required, we selected \texttt{set$(1,1)$dev$(3,3)$} for additional in-depth analysis due to the fact that it had two high contrast DIT peaks.
Here we show scanning transmission measurements for a number of the other devices.

\begin{table}[h]
    \centering
    \begin{tabularx}{0.85\textwidth}{ l l l X }
        \hline\hline \\
        Device & Wavelength (nm)\qquad\ & Quality factor\qquad\ & Comment \\
        \\
        \hline
        \\
        \texttt{set$(1,1)$dev$(2,1)$}\qquad\ & 735.60 &  4000 & Moderate DIT signal. \\
        \texttt{set$(1,1)$dev$(2,2)$} & 735.77 &  4000 & Moderate DIT signal. \\
        \texttt{set$(1,1)$dev$(3,3)$} & 735.68 &  3000 & Device from main text. \\
        \texttt{set$(1,1)$dev$(6,3)$} & 735.38 &  6000 & Not measured. \\
        \texttt{set$(1,1)$dev$(7,3)$} & 735.21 &  7000 & No DIT signal (likely poor coupling). \\
        \\
        \texttt{set$(2,2)$dev$(5,3)$} & 736.41 &  5000 & Strong DIT signal. \\
        \texttt{set$(2,2)$dev$(8,3)$} & 735.56 & 10000 & Not measured. \\
        \texttt{set$(2,2)$dev$(9,3)$} & 736.22 & 10000 & No DIT signal (likely poor coupling). \\
        \texttt{set$(2,2)$dev$(3,4)$} & 736.23 &  3000 & Strong DIT signal. \\
        \\
        \texttt{set$(2,3)$dev$(8,1)$} & 736.44 &  4000 & Moderate DIT signal. \\
        \texttt{set$(2,3)$dev$(5,2)$} & 736.19 &  4000 & Not measured. \\
        \texttt{set$(2,3)$dev$(6,4)$} & 735.34 & 10000 & Strong DIT signal. \\
        \\
        \hline\hline
    \end{tabularx}
    \caption{\textbf{Devices of interest for DIT measurements.}}
    \label{tab:si:dit_devices}
\end{table}

\begin{figure}[h]
    \centering
    \includegraphics{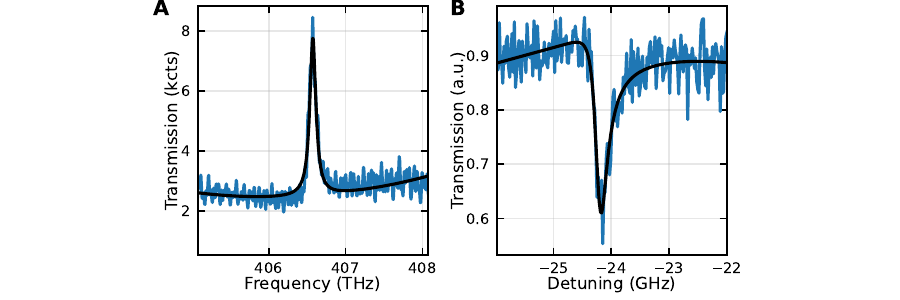}
    \caption{
        \textbf{Transmission measurements on \texttt{set(2,2)dev(5,3)}.}
        (\textbf{A}) Broadband transmission measurement with fit to a coupled-mode theory model.
        The cavity has a quality factor $Q = 4300$
        (\textbf{B}) Fit of a DIT line in the cavity to the input-output theory model.
        The estimated cooperativity is $C>(0.50\pm0.02)$, which is again a lower bound due to the unaccounted for thermalization of the SiV ground state orbitals.
        We did not include a full-range scan of this device as the wavemeter data could not be extracted for that measurement.
    }
    \label{fig:dit_p22c_dev05}
\end{figure}

\begin{figure}[h]
    \centering
    \includegraphics{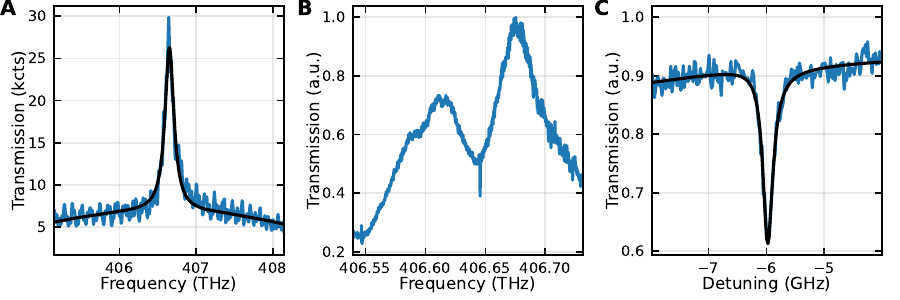}
    \caption{
        \textbf{Transmission measurements on \texttt{set(2,2)dev(3,4)}.}
        (\textbf{A}) Broadband transmission measurement with fit to a coupled-mode theory model.
        The cavity has a quality factor $Q = 2900$
        (\textbf{B}) Full-range laser scan over the resonance normalized to the input power.
        There remain Fabry-Perot oscillations from the microscope optics.
        There is a single strong DIT dip near the center of the resonance.
        (\textbf{C}) Fit of the DIT line.
        The estimated cooperativity is $C>(0.36\pm0.01)$, which is again a lower bound due to the unaccounted for thermalization of the SiV ground state orbitals.
    }
    \label{fig:dit_p22d_dev03}
\end{figure}

\begin{figure}[h]
    \centering
    \includegraphics{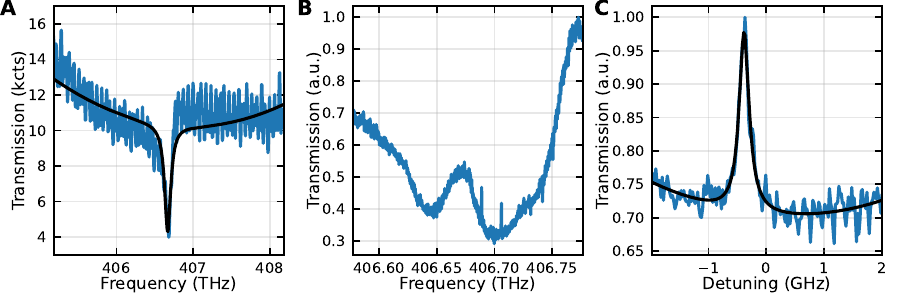}
    \caption{
        \textbf{Transmission measurements on \texttt{set(2,3)dev(8,1)}.}
        (\textbf{A}) Broadband transmission measurement with fit to a coupled-mode theory model.
        The cavity has a quality factor $Q = 4200$.
        Note that this is a thru device.
        The fit here assumes that there is no TM background, which complicates the analysis of the DIT signal.
        (\textbf{B}) Full-range laser scan over the resonance normalized to the input power.
        There remain Fabry-Perot oscillations from the microscope optics.
        Two strong DIT peaks are observed in the center of the cavity, there are also some detuned DIT peaks at higher frequencies.
        (\textbf{C}) Narrow-range DIT spectrum of the left peak with fit to input-output theory model.
        The fit in this case is complicated by the fact that the thru device DIT spectrum depends on the ratio between the coupling $\kappa_c$ and intrinsic $\kappa_i$ losses.
        Since we were unable to eliminate the background signal, it is not possible to obtain a high-confidence estimate of the cooperativity.
        Assuming there is no background enables a lower bound on $\kappa_c$, which in turn enables fits that typically lower bound the cooperativity.
        For this particular line we estimate $C > (0.45\pm0.01)$.
        We note that this is a lower bound both because of the uncertain coupling rate and because of thermalization of the orbitals is unaccounted for, and so this estimate is likely much lower than the real value.
    }
    \label{fig:dit_p23a_dev08}
\end{figure}

\clearpage
\subsection{Magnetic field mount}
\begin{figure}[h]
    \centering
    \includegraphics{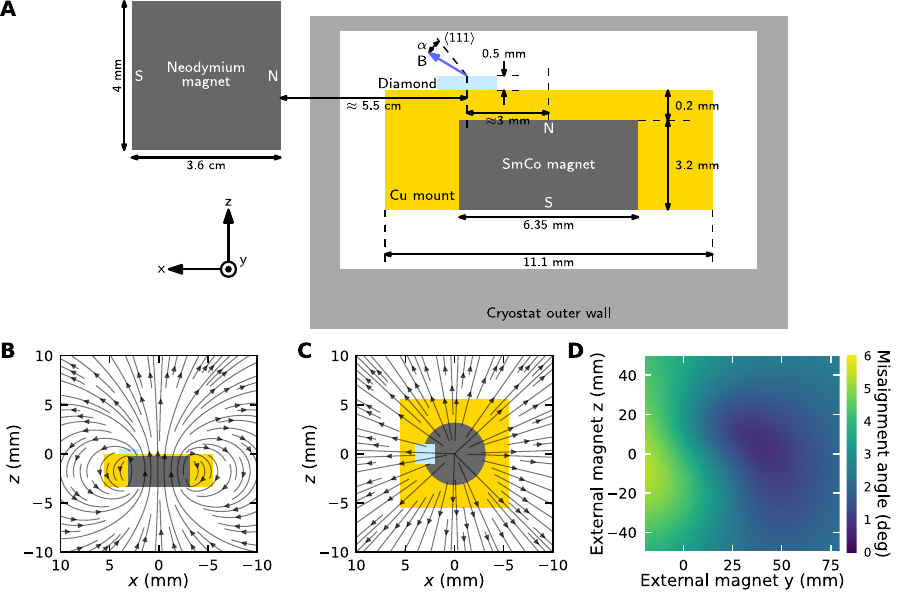}
    \caption{
        \textbf{Magnetic mount.}
        (\textbf{A}) Schematic of the magnetic mount with external neodymium mount.
        The drawing is not to scale.
        Field lines of the internal magnet (without the external Neodymium magnet) in the (\textbf{B}) $xz$ and (\textbf{C}) $xy$ planes.
        (\textbf{D}) Plot of the simulated misalignment angle $\alpha$ as a function of the external magnet position.
        The external magnet is assumed to be at $x=5.7$\,cm.
        The field strength is not significantly modified by the external magnet.
    }
    \label{fig:si:magnet}
\end{figure}

The cavity axis is aligned along $[110]$ during stamping which allows for the SiV orientations corresponding to $[\bar{1}11]$ and $[1\bar{1}1]$ to maximally couple to the TE mode of the cavity.
Both SiV 1 and 2 are expected to be one of these two orientations as their strong coupling would not be attainable otherwise.
We design the mount with an embedded samarium cobalt (SmCo) permanent magnet (diameter: $6.35$\,mm, thickness: $3.2$\,mm) as shown in Fig.~\ref{fig:si:magnet}.
The ideal position of the devices is at approximately $(x,y) = (3,0)$\,mm from the center of the mount.
There is considerable variation of the field angle over the 2-mm diamond and so we target alignment for only a single set of devices.

The sample is placed by hand and the position of the device of interest is approximately determined to be at $(x,y) = (3.014,0.190)$\,mm via wide-field imaging.
There is also a slight rotation of the diamond with respect to the mount on the order of $1^\circ$, however this is not taken into account for the simulations.
We simulate the field at the device of interest using \texttt{Magpylib}~\cite{ortner2020magpylib} and estimate a misalignment angle of $\alpha\approx3^\circ$ and a field strength of $|\vb B| \approx 0.26$\,T.

The external neodymium magnet (diameter: $4$\,cm, thickness: $3.6$\,cm) is mounted to a three-axis stage and is used to adjust the field angle further.
It produces considerably weaker fields on the order of 10\,mT due to the increased separation from the sample. 
We simulate the field at the devices as the magnet is swept in the $yz$ plane at its closest to the sample ($x\approx5.5$\,cm).
The final optimized position is centered roughly at $(y,z) = (40.3, 5.5)$\,mm and is relatively insensitive to the exact position within a few millimeters.
If placed optimally, we estimate a misalignment angle of $\alpha\approx0.8^\circ$ and a field strength of $|\vb B| \approx 0.25$\,T.

\clearpage
\subsection{Spin relaxation and optical transmission switching}

\subsubsection{Off-axis field}
While we know that SiV 1 and 2 are aligned along either $[\bar{1}11]$ and $[1\bar{1}1]$, it is not straightforwardly possible to distinguish between these orientations via PL measurements.
We are thus required to ``guess-and-check'' the field alignment.
Our initial placement of the sample aligned the field along $[\bar{1}11]$, however it was later determined that both SiV are instead aligned along $[1\bar{1}1]$.
Consequently, the field in the following measurements has an approximate misalignment angle $\alpha\approx 70^\circ$.
Simulation of the magnetic field (given the imperfect diamond placement) estimates the field strength to be about $0.24$\,T and with a roughly $65^\circ$ offset from $[1\bar{1}1]$.

\begin{figure}[h]
    \centering
    \includegraphics{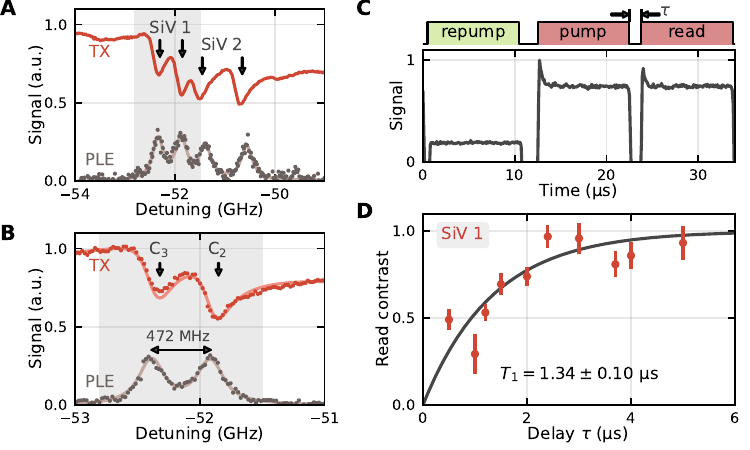}
    \caption{
        \textbf{Spin relaxation with misaligned field.}
        (\textbf{A}) Transmission and PLE signal from the device in a misaligned field ($\alpha\approx70^\circ$).
        SiV 1 and 2 are closer together due to random strain.
        (\textbf{B}) Transmission and PLE signal from SiV 1 after selectively ionizing SiV 2.
        (\textbf{C}) Representative spin pumping signal taken by resonantly exciting the $C_3$ transition of SiV 1.
        (\textbf{D}) Spin recovery curve as a function of time between the pump and read pulses.
    }
    \label{fig:si:spin_misaligned}
\end{figure}

The cryostat was cycled several times in this orientation.
Fig.~\ref{fig:si:spin_misaligned} shows the PLE and DIT spectra when SiV 1 and 2 are nearly overlapping due to stochastic strain induced during the cooldown.
The cavity is tuned such that the SiV are detuned at approximately $\Delta/2\pi \approx -50$\,GHz.
The measured splitting of the spin-conserving transitions is approximately 472\,MHz and 815\,MHz for SiV 1 and 2 respectively.
Given the expected field strength, such large splitting for SiV 2 immediately indicates that it is likely in the unaligned orientation (i.e.\ $[1\bar{1}1]$).
SiV 1, however, was indeterminate at this point.

We performed spin-relaxation measurements on SiV 1, resonantly driving the $C_3$ transition with 5--10\,nW (CW) to optically pump the SiV into $\ket{\uparrow}$ and readout the $\ket{\downarrow}$ population (Fig.~\ref{fig:si:spin_misaligned}C).
Note that these powers are approaching the SiV saturation and so the DIT signal begins to degrade.
Instead, these measurements capture the PLE signal.
The observed spin pumping times are less than 1\,\textmu s, suggesting the SiV is likely in the misaligned orientation (i.e.\ $[1\bar{1}1]$).
Fitting the full sequence of measurements yields a spin-relaxation time $T_1 = (1.34\pm0.10)$\,\textmu s (Fig.~\ref{fig:si:spin_misaligned}D).

\begin{figure}[h]
    \centering
    \includegraphics{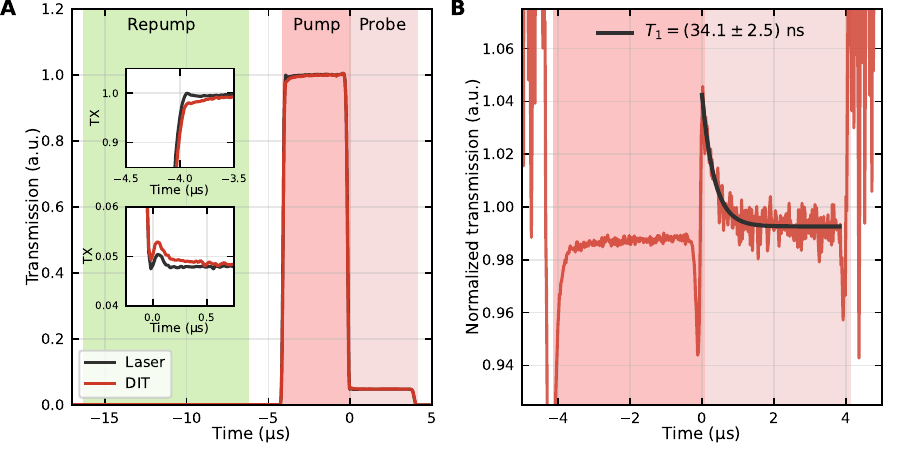}
    \caption{
        \textbf{Optical switching with misaligned field.}
        (\textbf{A}) Raw transmission signals through the cavity on resonance with SiV 1 (``DIT'') or off resonance (``Laser'').
        Inset plots shows the leading edge of the pump and probe pulses where the optical pumping occurs.
        (\textbf{B}) Normalized transmission (ratio of the resonant/non-resonant signals) with fit to the DIT dip recovery yielding the spin relaxation time.
    }
    \label{fig:si:pumping_misaligned}
\end{figure}

We also perform optical transmission switching measurements where the pump pulse is of similar power but the probe pulse is attenuated to approximately 10\% of the pump power which enables unsaturated DIT signal.
The raw signals when (non-)resonant with the $C_3$ transition of SiV 1 are shown in Fig.~\ref{fig:si:pumping_misaligned}A with their ratio given in Fig.~\ref{fig:si:pumping_misaligned}B.
We observe an exponential decay during the probe pulse consistent with the spin relaxing back into $\ket{\downarrow}$ and the DIT signal being restored.
The characteristic time corresponds to $0.34$\,\textmu s, which is roughly consistent with the approximately 1-\textmu s spin relaxation time measured previously.
The difference in value may be attributed to the fact that these measurements were performed on different cooldown cycles and so the field/strain have likely changed.

\subsubsection{Near-axis field}
The sample was then reoriented on the mount by a $180^\circ$ rotation.
The cavity was tuned onto near resonance with SiV 1 and 2 ($|\Delta|/2\pi < 10$\,GHz).
Simulation of the magnetic field estimates the field strength to be approximately $0.26$\,T with misalignment angle $\alpha\approx4^\circ$ due to imperfect placement of the sample.

The splitting of the spin-conserving transitions was estimated to be around 452\,MHz and 526\,MHz for SiV 1 and 2 respectively.
This enables optical resolution of the Purcell-broadened lines which are estimated to have line widths of 318\,MHz and 285\,MHz respectively.
Note that these line widths are consistent with the measured broadening at zero-field (main text, Fig.~4D).

\begin{figure}[h]
    \centering
    \includegraphics{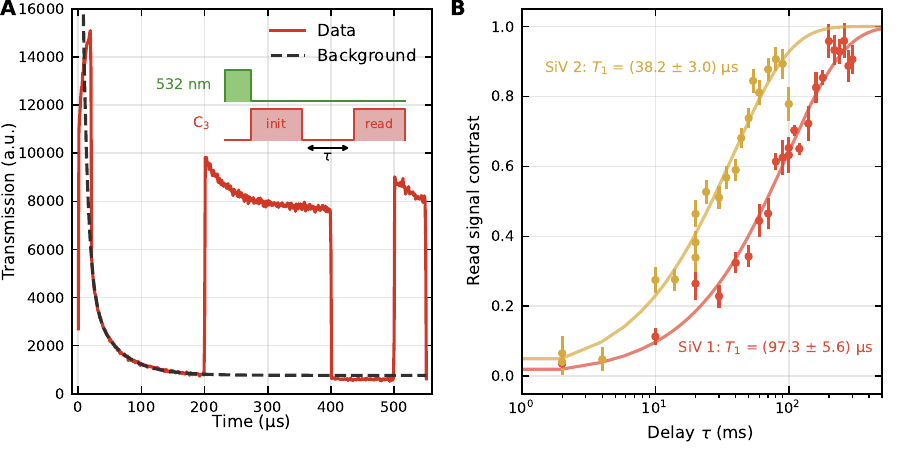}
    \caption{
        \textbf{Spin relaxation in un-optimized field ($\alpha\approx 4^\circ)$.}
        (\textbf{A}) Representative spin pumping signal for driving the $C_3$ transition of SiV 1.
        The transient decay signal from the pump step is of unknown origin.
        We remove it by fitting to a bi-exponential curve.
        (\textbf{B}) Spin relaxation curves for SiV 1 and 2.
    }
    \label{fig:si:spin_unoptimized}
\end{figure}

Similar spin-relaxation measurements are performed with slightly elevated CW powers of 10-15\,nW, also resonantly exciting the $C_3$ transition of either SiV 1 or 2 (Fig.~\ref{fig:si:spin_unoptimized}).
The slight increase in power is due to the balance between the increased cyclicity (which reduces the effective pumping rate) and the increased $T_1$ (which relaxes the pumping rate requirement).
We observe spin pumping times on the order of tens of microseconds and adjust the pulse sequence accordingly.
We note that the green repump AOM began to exhibit an unknown transient behavior which results in a weak background signal decaying over 100\,\textmu s after the end of the pulse.
To compensate, we increase the time separation between the green repump pulse and our pump/readout pulses.
We also remove the transient from the signal as background by fitting it to a two-rate exponential decay which qualitatively works well.
Fitting to the recovery signal as before, we obtain spin-relaxation times $T_1 = (97.3\pm5.6)$\,\textmu s and $T_1 = (38.2\pm3.0)$\,\textmu s for SiV 1 and SiV 2 respectively.

We perform similar transmission switching measurements as before with adjusted time scales for the longer $T_1$.
We only look at SiV 1 due to its longer $T_1$.
A simlar exponential decay is observed during the probe step with characteristic time scale of around $99$\,\textmu s, consistent with the measured $T_1$. 

\subsubsection{Field alignment}
The SiV $T_1$ and optical pumping time is expected to be highly sensitive to the misalignment angle $\alpha$ for near alignment.
We simulated the field angle as a function of the position of a large ``external'' neodymium magnet mounted to a three-axis stage outside of the cryostat chamber.
We roughly aign the external magnet to the simulated optimal position, then adjust using the translation stage to optimize the alignment.

The alignment process uses a spin-relaxation pulse sequence with fixed delay $\tau$.
Since the $T_1$ should increase rapidly as $\alpha\to0^\circ$, we expect the recovery of the PLE signal on the readout pulse to diminish as the field becomes more aligned (or increase as the misalignment angle increases).
We iterate on this process, increasing $\tau$ when the readout signal is reduced below the noise.
Our final alignment yields the spin-relaxation and transmission switching curves presented in the main text (Fig.~5).
The resulting spin-relaxation time for SiV 1 is measured to be $T_1=(419\pm18)$\,\textmu s, which is about four times larger than the un-optimized value.
Simulations indicate that the $\alpha<1^\circ$ at this position.
The final misalignment angle is limited by the comparatively weaker field of the external magnet ($B\approx 10$\,mT compared to $B\approx 100$\,mT from the internal magnet) and the high sensitivity of the applied field on our sample placement.

\clearpage
\subsection{Single-shot readout}
\begin{figure}[h]
    \centering
    \includegraphics{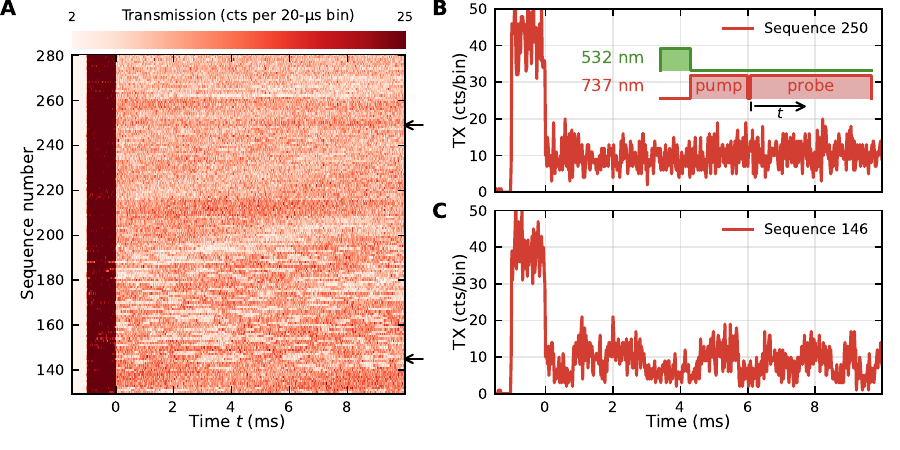}
    \caption{
        \textbf{Raw single-shot readout data.}
        (\textbf{A}) Transmission data for 150 consecutive single-shot measurement sequences (out of 1000 total in this measurement).
        The repump and pump steps occur at times $t<0$.
        The jumps appear between sequences 140--180 as blinking between two intensities.
        Elsewhere the signal appears as shot noise.
        (\textbf{B}) A representative sequence corresponding to non-resonant driving of the SiV (indicated by the arrow on (\textbf{A}).
        Inset shows the pulse sequence used in the measurements.
        (\textbf{C}) A representative sequence corresponding to visible quantum jumps of the spin.
    }
    \label{fig:si:single_shot}
\end{figure}

Single-shot readout measurements are performed using the same type of pump-probe pulse sequence used for the transmission switching measurements.
Counts from each cycle are recorded as a function of time into 20\,\textmu s bins.
Fig.~\ref{fig:si:single_shot}A shows 150 representative pulse sequences taken sequentially within a 1000-sequence measurement.

Due to transient effects in the AOMs (likely attributed to heating), the average power in the probe step monotonically decreases during the measurement sequence.
This, in effect, results in a different input power (and different average photon flux) during the measurement.
With respect to single-shot readout, this variation is prohibitive as states are discriminated by photon counts, which are assumed to occur at a constant rate in either state.
Additionally, there are slow oscillations in the signal which we attribute to vibrations of the sample stage due to the cryostat cycle.
These produce slow variation in the averaged transmitted signal which similarly degrades the single-shot measurement signal.
To deal with both of these effects, we manually post-select individual sequences for which the effect of the vibration is minimal, and then process the data in batches over which the power variation can be neglected.

Another notable feature of the  data is that the fluctuations due to the spin thermalization (i.e.\ single-shot readout) can be clearly observed to disappear and reappear at different points throughout the measurement.
It is not clear if this is due to spectral diffusion or ionization into a dark charge state.
However, the SiV tends to remain at approximately the same frequency over multiple measurement sequences lasting tens of minutes and so we can likely attribute such fast fluctuations to a measurement effect.
We hypothesize that the green repump pulse (which has a comparatively higher power of around 10\,\textmu W) is likely the source of the fluctuations, either by direct interaction with the SiV itself, or by affecting the charge environment via nearby defects.
Incidentally, this is likely the reason for the reduced spin-initialization fidelity in the spin-relaxation and transmission switching measurements.

\begin{figure}[h]
    \centering
    \includegraphics{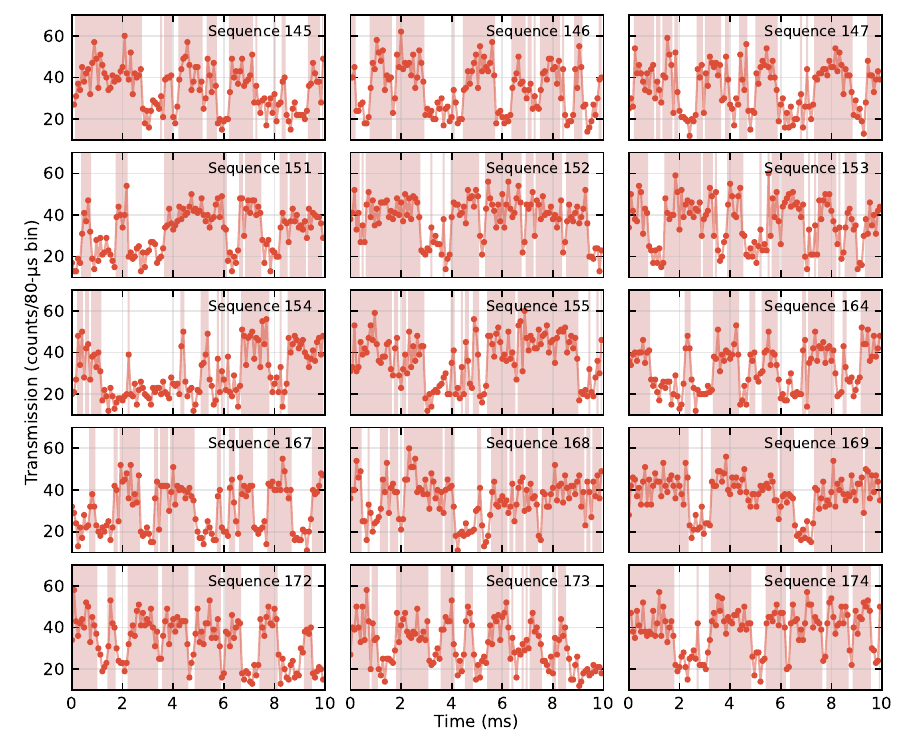}
    \caption{
        \textbf{Single-shot measurement sequences.}
        The 15 post-selected measurement sequences used for the main text.
        Projected states (indicated by the shading) are determined using the optimal threshold of 30\,cts/bin.
        Sequence 164 is shown in main text Fig.~5.
    }
    \label{fig:si:single_shot_jumps}
\end{figure}

We bin the data further during processing to 80-\textmu s windows to increase the signal-to-noise and improve the single-shot readout fidelity.
The 80-\textmu s bin window is sufficiently long to clearly distinguish the two spin states, while also being short enough to resolve the thermal fluctuations of the spin (recall $T_1 \approx 400$\,\textmu s which is five-times the bin size).
We examine 15 sequences from a 30-sequence long window (chosen as described above), which are plotted in Fig.~\ref{fig:si:single_shot_jumps}.
We bin the intensity distribution over all sequences and fit to a bimodal Poisson distribution 
\begin{equation*}
    f[k] = a\frac{\mu_\downarrow^k e^{-\mu_\downarrow}}{k!}
    +
    b\frac{\mu_\uparrow^k e^{-\mu_\uparrow}}{k!},
\end{equation*}
where $a$ and $b$ are arbitrary constants.
We obtain average count rates $\mu_\downarrow = (21.9\pm0.2)$\,cts/bin and $\mu_\uparrow = (41.3\pm0.2)$\,cts/bin for $\ket{\downarrow}$ and $\ket{\uparrow}$ respectively.
The single-shot readout fidelity is defined by~\cite{robledo2011fidelity}
\begin{equation}
    F = 1 - \frac{1}{2}\Big( 
        p(\downarrow|\uparrow) + p(\uparrow|\downarrow)
    \Big),
\end{equation}
where $p(\downarrow|\uparrow)$ is the conditional probability of measuring a given shot as $\ket{\downarrow}$ when the actual state is $\ket{\uparrow}$ (and vice versa for $p(\uparrow|\downarrow)$).
This depends on the measurement discrimination threshold $T$ which defines the count rate over which a given shot is assigned to $\ket{\uparrow}$ (i.e.\ if $k>T$ counts are detected the state is readout as $\ket{\uparrow}$, and if $k\leq T$ then the state is readout as $\ket{\downarrow}$).
In this scheme the probabilities can be expressed in terms of the cumulative distribution function (CDF) of the Poissonian intensity distribution as
\begin{align*}
    p(\downarrow|\uparrow) &= \mathrm{CDF}(T,\mu_\uparrow), \\
    p(\uparrow|\downarrow) &= 1 - \mathrm{CDF}(T,\mu_\downarrow).
\end{align*}
Uncertainty in the fidelity (due to imperfect knowledge of the rates $\mu_{\uparrow,\downarrow}$) is obtained as
\begin{equation*}
    \Delta F = \frac{1}{2} \sqrt{
        \Big( \pdv{p(\downarrow|\uparrow)}{\mu_\uparrow} \Big)^2\Delta\mu_\uparrow
        +
        \Big( \pdv{p(\uparrow|\downarrow)}{\mu_\downarrow} \Big)^2\Delta\mu_\downarrow
    },
\end{equation*}
where we use the fact that, for the Poisson distribution,
\begin{equation*}
    \pdv{\mathrm{CDF}(T,\mu)}{\mu} = \mathrm{CDF}(T-1,\mu) - \mathrm{CDF}(T,\mu) = -\frac{\mu^T e^{-\mu}}{T!}.
\end{equation*}
The fidelity is calculated as a function of the threshold $T$ to determine the optimal value of $F = (96.0\pm0.3)$\% with $T=30$.

After determining the optimal threshold we can then assign the shots to either $\ket{\uparrow}$ or $\ket{\downarrow}$ accordingly.
We can then examine the statistics of the thermal spin fluctuations.
For all measurement sequences we count the time between subsequent spin flip events and plot the resulting distribution.
The time between subsequent jumps $\tau$ is expected to follow an exponential distribution $p(\tau) = (1/T_1) \exp(-\tau/T_1)$, where the time constant is the spin relaxation time.
Note that we are unable to resolve time between jumps less than the 80-\textmu s bin window.
Fitting the data directly (neglecting $\tau<80$\,\textmu s), we obtain an estimate of the spin-relaxation time $T_1=(366\pm26)$\,\textmu s.
Notably, the imperfect fidelity and moderate noise in the transmission signal results in numerous ``false jumps'' wherein the shot noise crosses the threshold despite the spin not actually flipping.
Such false jumps will overwhelmingly occur as single, isolated shots and thus create additional signal in the lowest 80-\textmu s bin.
If we fit the binned data ignoring this first bin we obtain $T_1=(411\pm33)$\,\textmu s which is much closer to the previously measured $T_1$ from the spin-relaxation measurements.

\clearpage
\bibliography{main}